\documentclass[a4paper,12pt]{article}
\usepackage{latexsym}
\usepackage{euscript}
\usepackage{epsfig}
\usepackage{citesort}
\date{\today}
 \large\normalsize
\newcommand{\no}{\nonumber\\}

\def\tgb{\mbox{$\tan{\beta}~$}}
\def\bsg{$b\to s \gamma$~}
\def\eps{$\varepsilon$~}
\def\epspeps{$\varepsilon^{\prime}/\varepsilon$~}

\def\mch{$m_{\chi^{\pm}}$~}

\def\mgrav{$m_{3/2}$~}
\def\Ibanez{Iba\~{n}ez~}
\def\Munoz{Mu\~{n}oz~}
\newcommand{\BXcenu}{B\rightarrow X_c e \nu}
\newcommand{\mub}{\mu_b}
\newcommand{\mb}{m_b}
\newcommand{\alphas}{\alpha_s}
\newcommand{\alphae}{\alpha_e}
\newcommand{\BRg}{{\rm BR}(B\to X_s \gamma)}
\newcommand{\BR}{{\rm BR}}
\newcommand{\nl}{\nonumber \\}
\newcommand{\Bsg}{B\to X_s \gamma}

\def\beq{\begin{equation}}
\def\eeq{\end{equation}}
\def\bea{\begin{eqnarray}}
\def\eea{\end{eqnarray}}

\def\Frac#1#2{\frac{\displaystyle{#1}}{\displaystyle{#2}}}
\def\lsim{\raise0.3ex\hbox{$\;<$\kern-0.75em\raise-1.1ex\hbox{$\sim\;$}}}
\def\gsim{\raise0.3ex\hbox{$\;>$\kern-0.75em\raise-1.1ex\hbox{$\sim\;$}}}

\renewcommand{\O}{{\cal O}}
\def\npb#1#2#3{    {\it Nucl. Phys. }{\bf B #1} (#2) #3}

\def\plb#1#2#3{    {\it Phys. Lett. }{\bf B #1} (#2) #3}
\def\prd#1#2#3{    {\it Phys. Rev. }{\bf D #1} (#2) #3}

\def\prl#1#2#3{    {\it Phys. Rev. Lett. }{\bf #1} (#2) #3}

\begin{document}

%\draft

\renewcommand{\thefootnote}{\fnsymbol{footnote}}

\begin{titlepage}
\begin{flushright}
FTUAM-00-11\\
hep-ph/0005303
\end{flushright}
\vspace{1cm}
\begin{center}
{\Large {\bf The $b\to s\gamma$ decay 
in SUSY models with\\[0.2cm] non--universal $A$--terms}}
\vspace{1cm}  

{E. Gabrielli$^{1}$\footnote{emidio.gabrielli@cern.ch}, S.
Khalil$^{1,2}$\footnote{shaaban.khalil@uam.es}, and E.
Torrente-Lujan$^{1}$\footnote{e.torrente@cern.ch} 
}
\vspace{0.5cm}
\end{center}
\begin{center}
{\small
{\it $^1$ Dept. de F\'{\i}sica
Te\'orica C-XI, Universidad Aut\'onoma de Madrid,\\
Cantoblanco, 28049 Madrid, Spain. \\
\vspace*{2mm}
\it $^2$ Ain Shams University, Faculty of Science, Cairo
11566, Egypt.}
}
\end{center}

\bigskip
\begin{center}
{\bf Abstract}\\
\end{center}

We analyse the predictions for the inclusive branching ratio for $\Bsg$ 
in a class of string-inspired SUSY models with 
non--universal soft-breaking $A$--terms. 
These models are particularly interesting since the
non--universality of the $A$--terms plays an important role 
in providing new significant 
contributions to CP violation effects in kaon physics while respecting 
the severe bounds on electric dipole moments. 
We show that \bsg do not severely constrain the non--universality 
of these models. In particular, at low \tgb $(\tan\beta\simeq 2)$,
our predictions are close to the universal case. For large \tgb 
($\tan\beta \gsim 15$),the effect of non--universality is enhanced and 
stronger constraints hold.
We find that the parameter regions which are important for generating 
sizeable contribution to \epspeps, of order $2 \times 10^{-3}$, are not 
excluded.
\end{titlepage}

\newpage

\baselineskip 20pt

%%%%%%%%%%%%%%%%%%%%%%%%%%%%%%%%%%%%%%%%%%%%%%%%%%%%%%%%%%%%%%%%%%%%%%%%%%%%%
\section{ Introduction}
\label{sec:intro}
\par
Recently there has been a growing interest concerning supersymmetric 
models with non--universal soft--breaking terms 
\cite{nonuniv,abel,khalil1,khalil2,vives}.
The main theoretical 
reason is that some string-inspired models naturally favour 
SUSY models with non--universality in the soft-breaking sector
\cite{ibanezlust,kapl,ibanez1,ibanez2}. 
Within this class of string-inspired models particularly interesting are 
that ones with non--universal $A$--terms \cite{khalil1,khalil2}. 
This is mainly due to the 
relevant role they can play in solving the SUSY CP problem,
especially in the light of the recent experimental results on 
the direct CP violation parameter \epspeps. 

The new measurements of \epspeps at KTeV \cite{CP1} and 
NA48 \cite{CP2}
lead to a world average 
of Re \epspeps =$(21.4\pm 4.0)\times 10^{-4}$~\cite{CP3}.
This result is higher than the Standard Model (SM) predictions \cite{epsp1}, 
opening the way to the interpretation that it may be a signal of 
new physics beyond the SM.
Clearly it is still premature to claim that this is a genuine 
new physics effect, mainly due to the large theoretical uncertainties 
in the non-perturbative hadronic sector which affect the SM
predictions for \epspeps \cite{epsp2}.
However if one accepts the point of view that new CP sources are needed 
in order to obtain large values for \epspeps, one may wonder if 
the minimal supersymmetric extension of the SM (MSSM)
can help in solving this problem. 
The answer is no,
mainly due to the assumption of universal boundary conditions
of the soft-breaking terms \cite{khalil1,GG,masieromur,demir}.
Moreover we stress that without new flavor structure beyond the usual 
Yukawa couplings, general SUSY models with phases of the
soft terms of order $\O(1)$ (but 
with a vanishing CKM phase $\delta_{\mathrm{CKM}}=0$)
can not give a sizeable contribution to the CP violating processes
\cite{abel,khalil1,demir,barr}.

The impact of the new flavor structure in the non--universality of
the $A$--terms have been studied in 
Refs.~\cite{nonuniv,abel,khalil1,khalil2,vives}.
In these works it was emphasized that the non-degenerate 
$A$--terms can generate the
experimentally observed CP violation $\varepsilon$ and
$\varepsilon'/\varepsilon$ even with a vanishing 
$\delta_{CKM}$, i.e., fully supersymmetric CP violation
in the kaon system is possible in a class of models with non--universal
$A$--terms.
This effect can be simply understood 
by making use of the mass-insertion approximation \cite{GGMS} :
the non-degenerate $A$--terms enhance the gluino contributions 
to \eps and \epspeps through 
enhancing the imaginary parts of the L-R mass insertions 
$\mathrm{Im}(\delta^d_{LR})_{12}$ and $\mathrm{Im}(\delta^d_{RL})_{12}$~
\cite{khalil1}.

It is well known that the experimental 
\bsg constraints cause a dramatic reduction 
of the allowed
parameter space in case of universal soft terms \cite{bsgSUSY,BBMR}. 
Hence one may
worry if these constraints would be even more severe in the case of the
non-degenerate $A$--terms.
However a complete analysis of the \bsg constraints has not been considered
in Refs.~\cite{khalil1,khalil2,vives}.
In particular in these works it was roughly checked, by using the results
of Ref.~\cite{GGMS}, that the \bsg constraints are satisfied, namely 
$(\delta_{LR}^d)_{23} \leq 1.6\times 10^{-2}$ and
$(\delta_{LL}^d)_{23} \leq 8.2$.
Note that these constraints are obtained by assuming
that the gluino amplitude is the
dominant contribution to \bsg (dominant also with respect to the SM). 

Although the gluino contribution to $b\to s\gamma$ 
decay is usually very small in the universal case, being
proportional to the mass insertions $(\delta_{LR}^d)_{23}$, 
we could expect that
this contribution may be enhanced by the non--universality of the $A$--terms.
However the non--universal $A$--terms could also give large contributions to 
the chargino amplitude through the $(\delta_{LR}^u)_{23}$.
Even though the non-degenerate $A$--terms
enhance the gluino contribution, one can not a priori expect that
it will be the dominant effect. 
Therefore a careful analysis of the \bsg predictions, including
the full SUSY contributions, is necessary in this scenario.

The purpose of this work is to perform a complete 
analysis of the \bsg constraints for the SUSY models 
with non--universal $A$--terms studied in 
Refs.~\cite{khalil1,khalil2,vives}.
Indeed we will show that our results are different
from the naive expectations based only on the gluino-dominance approximation.
\par

The paper is organized as follows. In section 2 we present two
models with non--universal $A$--terms that have recently been considered 
for solving the SUSY CP problem. These two models are based
on weakly coupled heterotic string and type I string theories respectively.
In section 3 we present formulas for the total branching ratio taking into 
account the QCD corrections and we discuss the different SUSY contributions
to the \bsg decay in these models. Our numerical results for the predictions 
of the \bsg branching ratio are presented in section 4.
The last section is devoted to conclusions. Finally, various formulas are 
summarized in the appendix.
\newpage
%%%%%%%%%%%%%%%%%%%%%%%%%%%%%%%%%%%%%%%%%%%%%%%%%%%%%%%%%%%%%%%%%%%%%%%%%%%%%
\section{String inspired models with non-degenerate\\
 $A$--terms}
\label{sec:models}
\par
In this work we consider the class of string inspired model which
has been recently studied in 
Refs.~\cite{khalil1,khalil2,vives}. In
this class of models, the trilinear $A$--terms of the soft SUSY
breaking are non--universal. It was shown that this
non--universality among the $A$--terms plays an important role on
CP violating processes. In particular, it has been shown that
non-degenerate $A$-parameters can generate the experimentally
observed CP violation $\varepsilon$ and $\varepsilon'/\varepsilon$
even with a vanishing $\delta_{\mathrm{CKM}}$.
\par

Here we consider two models for non-degenerate $A$--terms. The
first model (model A) is based on weakly coupled heterotic strings,
where the dilaton and the moduli fields contribute to SUSY
breaking~\cite{ibanez1}. The second model (model B) is based on type
I string theory where the gauge group $SU(3) \times U(1)_Y$ is
originated from the $9$ brane and the gauge group $SU(2)$ is
originated from one of the $5$ branes~\cite{ibanez2}. 

In order to fix the conventions, the following Lagrangian
${\cal L}_{SB}$ for the soft-breaking terms is assumed 
\beq 
-{\cal
L}_{SB}=\frac{1}{6}h_{ijk} \phi_i\phi_j\phi_k + \frac{1}{2} (\mu
B)^{ij}\phi_i\phi_j+\frac{1}{2}(m^2)^j_i \phi^{\star
i}\phi_j+\frac{1}{2}M_a\lambda_a\lambda_a + h.c. 
\label{Lsoft}
\eeq 
where the $\phi_i$ are the scalar parts of the chiral
superfields $\Phi_i$ and $\lambda_a$ are the gaugino fields. In
the notation for the trilinear couplings, the $A$--terms
are defined as $h_{ijk}=Y_{ijk}A_{ijk}$ (indices not summed)
where $Y_{ijk}$ are the corresponding Yukawa couplings.
\subsection{Model A}
We start with the weakly coupled string-inspired supergravity
theory.
In this class of models, it is assumed that the superpotential of
the dilaton ($S$) and moduli ($T$) fields is
generated by some non-perturbative mechanism and 
the $F$-terms of $S$ and $T$ contribute to the SUSY breaking. 
Then one can parametrize the $F$-terms as~\cite{ibanez1}
\beq
F^S = \sqrt{3} m_{3/2} (S+S^*) \sin\theta,\hspace{0.75cm} F^T
=m_{3/2} (T+T^*) \cos\theta .
\eeq
Here $m_{3/2}$ is the gravitino mass, $n_i$ is the modular weight
and $\tan \theta$ corresponds to the ratio between the $F$-terms of $S$
and $T$. 
In this framework, the soft scalar masses $m_i$ and the gaugino masses 
$M_a$ are given by~\cite{ibanez1}
\begin{eqnarray}
m^2_i &=& m^2_{3/2}(1 + n_i \cos^2\theta), \label{scalar}\\ M_a
&=& \sqrt{3} m_{3/2} \sin\theta .\label{gaugino}
\end{eqnarray}
The  $A^{u,d}$-terms are  written as
\begin{eqnarray}
(A^{u,d})_{ij} &=& - \sqrt{3} m_{3/2} \sin\theta- m_{3/2}
\cos\theta (3 + n_i + n_j + n_{H_{u,d}}), \label{trilinear}
\end{eqnarray}
where $n_{i,j,k}$ are the modular weights of the fields
that are coupled by this $A$--term. As shown in
Eqs.(\ref{scalar}-\ref{trilinear}), the values of the soft SUSY
breaking parameters depend on the modular weight of the matter
fields. These modular weights $n_i$ are negative integers, and their
`natural' values (in case of $Z_N$ orbifolds) are 
$-1,-2$, and $-3$~ \cite{ibanezlust}.
If we assign $n_i=-1$ for the third family and $n_i=-2$ 
for the first and second
families (we also assume that $n_{H_1}=-1$ and $n_{H_2}=-2$). Note that 
with this choice of modular weights we have $m_{H_2}^2 < m_{H_1}^2$ 
which is favored for the electroweak breaking (EW) and all the squark mass 
matrices are equal. Also we find the following texture for the $A$-parameter 
matrix at the string scale
\begin{equation}
A^{u,d} = \left (
\begin{array}{ccc}
x_{u,d} & x_{u,d} & y_{u,d}\\
x_{u,d} & x_{u,d} & y_{u,d} \\
y_{u,d} & y_{u,d} & z_{u,d}
\end{array}
\right),
\label{AtermA}
\end{equation}
where
\begin{eqnarray}
x_u&=& m_{3/2}(-\sqrt{3} \sin\theta + 3  \cos\theta),\\
x_d&=&y_u= m_{3/2}(-\sqrt{3} \sin\theta + 2  \cos\theta),\\
y_d&=&z_u= m_{3/2}(-\sqrt{3} \sin\theta + \cos\theta),\\
z_d&=&-\sqrt{3}m_{3/2}\sin\theta.
\end{eqnarray}

By fixing the value of $\tan \beta$ we can determine the values of
$\mu$ and $B$ from the radiatively EW breaking conditions. 
Then all the SUSY particle spectrum is completely determined in terms of
$m_{3/2}$ and $\theta$. The non--universality of this model is
parameterized by the angle $\theta$ and the value $\theta =\pi/2$
corresponds to the universal limit for the soft terms. In order to
avoid negative mass squared in the scalar masses we restrict
ourselves to the case with $\cos^2 \theta < 1/2$. Such
restriction on $\theta$ makes the non--universality in the whole
soft SUSY breaking terms very limited. However, as shown in
\cite{khalil1,khalil2}, this small range of variation for the
non--universality is enough to generate sizeable SUSY CP violations
in K system. We emphasize that choosing modular
weights different from those assigned above, the allowed range of the 
soft SUSY breaking terms do not essentially change.  For instance, if we
assign $n_i=-3$ for the first family instead of $-2$, it may
appear that the non--universality among the entries of the
$A$--terms is enhanced. However in this case the angle $\theta$ is
more constrained than before ($\cos^2 \theta < 1/3$). 
We have checked that different choices for $n_i$ do not significantly 
affect our results for the \bsg branching ratio.

\subsection{Model B}
As mentioned in the introduction, this model is based on type I
string theory. Like model A, this is a good candidate for
generating sizeable SUSY CP violations. Recently, there has been
considerable interest in studying the phenomenological
implications of this class of models \cite{type1}. In type I string theory, non
universality in the scalar masses, $A$--terms and gaugino masses
can be naturally obtained~\cite{ibanez2}. Type I models contain
either 9 branes and three types of $5_i (i=1,2,3)$ branes or $7_i$
branes and 3 branes. From the phenomenological point of view there
is no difference between these two scenarios. Here we consider the
same model used in Ref.~\cite{vives}, where the gauge group
$SU(3)_C \times U(1)_Y$ is associated with 9 brane while $SU(2)_L$
is associated with $5_1$ brane.

If SUSY breaking is analysed, as in model A, 
in terms of the vevs of the dilaton and moduli fields \cite{ibanez2}
\beq
F^S = \sqrt{3} m_{3/2} (S+S^*) \sin\theta,\hspace{0.75cm} F^{T_i}
=m_{3/2} (T_i+T_i^*) \Theta_i \cos\theta~,
\eeq
where the angle $\theta$ and the parameter $\Theta_i$ with 
$\sum_i \left|\Theta_i\right|^2=1$, just parametrize the direction of the
goldstino in the $S$ and $T_i$ fields space .
Within this framework, the gaugino masses are~\cite{ibanez2} 
\bea
M_1 &=& M_3 = \sqrt{3} m_{3/2} \sin\theta ,\\ M_2 &=& \sqrt{3}
m_{3/2} \Theta_1 \cos \theta .\label{m2} 
\label{gauginoB} 
\eea 
In this case the quark doublets and the Higgs fields are assigned to
the open string which spans between the $5_1$ and $9$ branes.
While the quark singlets correspond to the open string which starts
and ends on the $9$ brane, such open string includes three sectors
which correspond to the three complex compact dimensions. If we
assign the quark singlets to different sectors we obtain
non--universal $A$--terms. It turns out that in this model the
trilinear couplings $A^u$ and $A^d$ are given
by~\cite{ibanez2,vives}
\begin{equation}
A^u=A^d = \left (
\begin{array}{ccc}
x & y & z\\
x & y & z \\
x & y & z
\end{array}
\right),
\label{AtermB1}
\end{equation}
where
\begin{eqnarray}
x &=& - \sqrt{3} m_{3/2}\left(\sin\theta + (\Theta_1 - \Theta_3) \cos\theta
\right),\\
y &=& - \sqrt{3} m_{3/2}\left(\sin\theta + (\Theta_1 - \Theta_2) \cos\theta
\right),\\
z &=& - \sqrt{3} m_{3/2} \sin\theta.
\label{AtermB2}
\end{eqnarray}
The soft scalar masses for quark-doublets and Higgs fields
$(m^2_L)$, and the quark-singlets $(m^2_{R_i})$ are given by 
\bea
m^2_L &=&  m_{3/2}^2 \left( 1- \frac{3}{2} (1-\Theta_1^2) \cos^2
\theta\right) ,\\ m^2_{R_i} &=&  m_{3/2}^2 \left( 1- 3 \Theta_i^2 \cos^2
\theta\right), 
\label{scalarB} 
\eea 
where $i$ refers to the three families. 
For $\Theta_{i} = 1/\sqrt{3}$ the $A$--terms and the
scalar masses are universal while the gaugino masses could be
non--universal. The universal gaugino masses are obtained at
$\theta=\pi/6$.
\\

It is worth mentioning that in these models (A and B) the gaugino
masses, the $A$--terms, and the $\mu$--term are in general complex.
However, by using $R$-rotation we can make the gaugino masses real
and we end up, in addition to the phase of $\mu$, with the phases
of the $A$--terms. The phase of $\mu$ is severely constrained by
the electric dipole moment (EDM) of the electron and the
neutron~\cite{barr}, while the phases of the $A$--terms are
essentially unconstrained. Thus one can set the phase of $\mu$ to
zero, as done in Ref.~\cite{khalil1,khalil2,vives}. Moreover it has been
shown that the phases of the $A$--terms can lead to sizeable
supersymmetric contribution to CP observables, in particular on
the direct CP violation $\varepsilon'/\varepsilon$. However, since
the total branching ratio $b \to s \gamma$ decay is a CP conserving 
observable, this should not be very effective in 
constraining the phases of the SUSY
soft-breaking terms. For this reason we have made in our analysis the
simplifying assumption to set to zero all the phases.

\subsection{Yukawa textures}
As emphasized in Refs.~\cite{khalil1,khalil2,vives}, in
models with non-degenerate $A$--terms we have to fix the Yukawa
matrices to completely specify the model. In fact, with universal
$A$--terms the textures of the Yukawa matrices at GUT scale affect
the physics at EW scale only through the quark masses and usual
CKM matrix, since the extra parameters contained in the Yukawa
matrices can be eliminated by unitary fields transformations. This
is no longer true with non-degenerate $A$--terms since in the
scalar potential the $A$--terms enter through the tensorial product
$(Y_q^A)_{ij}=(Y_q)_{ij}(A_q)_{ij}$. Thus the diagonalization of
$Y_q$  can not be done simultaneously with $Y^A_q$ (unlike the
universal case). Thus in the models with non--universal $A$--terms,
some extra degrees of freedom (in addition to the quark masses and
CKM matrix) contained in the Yukawa matrices become observable.
Hence, the analysis of the non-degenerate $A$--terms could shed
some light on the favoured Yukawa textures.
For instance in Ref.~\cite{khalil2}, using a symmetric Yukawa texture with
a symmetric $A$--terms, an accidental cancellation between the
different SUSY contributions to $\varepsilon'/\varepsilon$ was
found, cancellation which leads to a very small value for
$\varepsilon'/\varepsilon$. On the contrary, by using asymmetric
Yukawa matrices with symmetric $A$--terms, this cancellation does
not occur and it is found that $\varepsilon'/\varepsilon$ can be
easily of the order of the KTeV result.

Here we show  two realistic
 examples of  Yukawa matrix textures that
have already been used in Refs.~\cite{khalil1,khalil2,vives}. In the
first one we have the following symmetric  Yukawa matrices
\begin{eqnarray}
\begin{array}{lr}
Y^d= y^b \left(\begin{array}{ccc}\!\! 0\!\!&\!\! V_{12}\Frac{m_s}{m_b}~~
\!\!&
\!\!V_{13} \\  V_{12}\Frac{m_s}{m_b} & \Frac{m_s}{m_b} &  V_{23} \\
V_{13} &  V_{23} & 1 \end{array} \right)~,~  &
Y^u= y^t \left(\begin{array}{ccc}\!\! 0\!\! &\!\! 0\!\! &\!\!
V_{13}
\\ 0 & \Frac{m_c}{m_t} & 0 \\  V_{13} & 0 & 1 \end{array} \right),
\label{Yuk1}
\end{array}
\end{eqnarray}
where $y^{b,t}$ are the Yukawa couplings of the bottom and top respectively,
and $V$ is the CKM matrix.
The second example is based on the assumption that the CKM mixing
matrix originates from the down Yukawa couplings and that the Yukawa matrices
are hermitian.
\beq
Y^u=\frac{1}{v\cos{\beta}} {\rm diag}\left(
m_u,m_c,m_t\right)~,~~
Y^d=\frac{1}{v\sin{\beta}}  V^{\dagger} \cdot {\rm diag }
\left(m_d, m_s, m_b\right) \cdot  V
\label{Yuk2}
\eeq
Although the analysis of the CP violation  is quite sensitive to
the specific Yukawa matrix, we found that the branching ratio of $b \to s \gamma$
does not essentially depend on it. We checked this property
by using different Yukawa textures (the two examples presented here and others).
We will only present through all the paper the results 
concerning the second example just as a representative case.

Now we present the general expressions for the squark mass matrices 
in the non-universal case in the SCKM basis,
in this basis  the unitary matrices
$S^{U_{R,L}}$ and $S^{D_{R,L}}$ (obtained by a superfield rotation) 
are chosen to diagonalize the up-- and down-- 
Yukawa couplings $Y^{u,d}$ 
\beq
m_U=\frac{v \sin{\beta}}{\sqrt{2}}S_{U_R} (Y^{u,d})^T S_{U_L}^{\dag},~~
m_D=\frac{v \cos{\beta}}{\sqrt{2}}S_{D_R} (Y^{u,d})^T S_{D_L}^{\dag}
\eeq
where $T$ stands for the transpose and
$m_{U,D}$ are the diagonal up-- and down--quark mass matrices respectively.
In this basis the up and down squark mass matrices at low energy
(respectively $M^2_{\tilde{u}}$ and $M^2_{\tilde{d}}$) are given by 
\cite{misiak1}
\bea
\begin{array}{lr}
M^2_{\tilde{u},\tilde{d}}=
\left(\begin{array}{cc}
\left(M^2_{\tilde{u},\tilde{d}}\right)_{LL}
&\left(M^2_{\tilde{u},\tilde{d}}\right)_{LR}
\\
\left(M^2_{\tilde{u},\tilde{d}}\right)_{RL}
&\left(M^2_{\tilde{u},\tilde{d}}\right)_{RR}
\end{array} \right),
\label{sqmass}
\end{array}
\eea
where for the up-sector
\bea
\left(M^2_{\tilde{u}}\right)_{LL}&=&S_{U_L}M^2_{\tilde{Q}}S_{U_L}^{\dag}
+m_U^2 +\frac{m_Z^2}{6}(3-4\sin^2{\theta_W})\cos{2\beta},\no
\left(M^2_{\tilde{u}}\right)_{RR}&=&S_{U_R}(M^2_{\tilde{u}^c})^T S_{U_R}^{\dag}
+m_U^2 +\frac{2 m_Z^2}{3}\sin^2{\theta_W}\cos{2\beta},\no
\left(M^2_{\tilde{u}}\right)_{LR}&=&\left(M^2_{\tilde{u}}\right)_{RL}^{\dag}=
\mu~m_U~{\rm cot}{\beta}+\frac{v\sin{\beta}}{\sqrt{2}}
S_{U_L}Y^{A \star}_u S_{U_R}^{\dag}~,
\label{sqmass1}
\eea
and for the down-sector
\bea
\left(M^2_{\tilde{d}}\right)_{LL}&=&S_{D_L}M^2_{\tilde{Q}}S_{D_L}^{\dag}
+m_D^2 -\frac{m_Z^2}{6}(3-2\sin^2{\theta_W})\cos{2\beta},\no
\left(M^2_{\tilde{d}}\right)_{RR}&=&S_{D_R}(M^2_{\tilde{d}^c})^T S_{D_R}^{\dag}
+m_D^2 +\frac{2 m_Z^2}{3}\sin^2{\theta_W}\cos{2\beta},\no
\left(M^2_{\tilde{d}}\right)_{LR}&=&\left(M^2_{\tilde{d}}\right)_{RL}^{\dag}=
\mu~m_D \tan{\beta}+\frac{v\cos{\beta}}{\sqrt{2}}
S_{D_L}Y^{A \star}_d S_{D_R}^{\dag}
\label{sqmass2}
\eea
where $M^2_{\tilde{Q}}$ and $M^2_{\tilde{u}^c,\tilde{d}^c}$ are the
soft-breaking ($3\times 3$) mass matrices
for the squark doublet and singlets respectively.
Our convention for the sign of the $\mu$--term is 
opposite to the same one of Ref.~\cite{misiak1}.
Note that the matrices $S_{U,D}$, unlike in the universal case,
can not be re--absorbed in the definition of diagonal Yukawa couplings.

\section{The \bsg decay in SUSY models}
\label{sec:results}
\par
In this section we analyse the \bsg decay in SUSY models with non--universal
$A$--terms.
As pointed out previously, these models are particularly interesting
because they can give sizeable contributions to the CP violating processes
through their large contributions to the mass insertions 
$(\delta_{LR}^{u,d})_{ij}$  or $(\delta_{RL}^{u,d})_{ij}$.
For this reason one may expect that large effects can be also induced
in the processes mediated by the dipole-magnetic operators, such as 
the rare decay \bsg.
Indeed, if the $\delta_{LR}^{u,d}$ are large enough,
then the SUSY contributions 
to these operators are enhanced since the typical chiral 
suppression is removed by the insertion of the internal gaugino mass,
thus allowing for a competition with the chiral-suppressed SM amplitude.

Let us start with the experimental results.
The most recent result reported by CLEO collaboration 
for the total (inclusive) B meson branching ratio $\Bsg$ is~ \cite{CLEO}
\beq
{\rm BR}(B\to X_s\gamma)=(3.15\pm 0.35\pm 0.32\pm 0.26)\times 10^{-4}
\label{bsgCLEO1}
\eeq
where the first error is statistical, the second systematic, and the
third one accounts for model dependence. From this result the following
bounds (each of them at 95\% C.L.) are obtained
\beq
2.0\times 10^{-4} < 
{\rm BR}(B\to X_s\gamma)< 4.5 \times 10^{-4}.
\label{bsgCLEO2}
\eeq
In addition the ALEPH collaboration at LEP reported a compatible measurement
of the corresponding branching ratio for b hadrons at the Z resonance
\cite{ALEPH}.

The starting point for the theoretical study of \bsg decay 
is given by the effective Hamiltonian
\beq
H_{eff}=-\frac{4G_F}{\sqrt{2}}V_{32}^{\star}V_{33}\sum_{i=1}^{8} 
C_i(\mu_b) Q_i(\mu_b)
\label{Heff}
\eeq
where the complete basis of operators in the SM can be found in 
Ref.~\cite{bsgNLO}.
Recently the main theoretical uncertainties present in the previous leading
order (LO) SM calculations have been reduced by including the 
NLO corrections to the \bsg decay, through
the calculation of the three-loop anomalous dimension matrix of the 
effective theory \cite{bsgNLO}.
The relevant SUSY contributions to the effective Hamiltonian in Eq.(\ref{Heff})
affect only the $Q_7$ and $Q_8$ operators, 
the expression for these operators are given (in the usual notation) by
\bea Q_7&=&\frac{e}{16 \pi^2}m_b\left(
\bar{s}_L\sigma^{\mu\nu} b_R\right) F_{\mu\nu}~,
\no
Q_8&=&\frac{g_s}{16
\pi^2}m_b\left( \bar{s}_L\sigma^{\mu\nu} T^a b_R\right) G^a_{\mu\nu}~.
\label{operators}
\eea
The Wilson coefficients $C_i(\mu)$ are evaluated at the renormalization scale
$\mu_b\simeq O(m_b)$  by including the
NLO corrections \cite{bsgNLO}.
They can be formally decomposed  as follows
\beq
C_i(\mu)=C_i^{(0)}(\mu)+\frac{\alphas(\mu)}{4\pi}C_i^{(1)}(\mu) +
\O(\alphas^2). 
\eeq
where $C_i^{(0)}$ and $C_i^{(1)}$ stand for the LO and NLO order
respectively.
Finally the branching ratio $\BRg$, conventionally normalized to the
semileptonic branching ratio $\BR^{exp}(\BXcenu)=(10.4\pm0.4)\%$
~\cite{PDG}, is given by~\cite{bsgNLO} 
\bea
{\rm BR}^{\rm NLO}(B\to X_s\gamma)
&=&\BR^{exp}(\BXcenu ) \frac{|V_{32}^{*} V_{33}|^2}{|V_{23}|^2}
\frac{6 \alphae}{\pi g(z) k(z)}\left(1-\frac{8}{3}\frac{\alphas
(\mb)}{\pi}\right) \no
&\times& \left(|D|^2+A\right)(1+\delta_{np})~,
\label{BRSM}
\eea
with
\bea
D&=& C_7^{(0)}(\mub)+\frac{\alphas (\mub)}{4\pi} \left(C_7^{(1)}(\mub)+
\sum_{i=1}^{8}C_i^{(0)}(\mub) \left[r_i(z)+\gamma_{i7}^{(0)}
\log{\frac{m_b}{\mub}}\right]\right), \nl 
A&=& \left( e^{-\alphas(\mub)
\log{\delta (7+2\log{\delta})/3\pi}}-1\right) |C_7^{(0)}(\mub )|^2 
+\frac{\alphas (\mub)}{\pi}\sum_{i\le j=1}^{8}C_i^{(0)}(\mub )
C_j^{(0)}(\mub ) f_{ij}(\delta)~,\nonumber
\eea 
where $z=m_c^2/m_b^2$. The expressions for $C_i^{(0)}$, $C_i^{(1)}$, and
the anomalous dimension matrix $\gamma$,
together with the functions $g(z)$, $k(z)$, $r_i(z)$ and 
$f_{ij}(\delta)$, can be found in
Ref.~\cite{bsgNLO}.  The term $\delta_{np}$ (of order a few
percent) includes the non-perturbative $1/\mb$ \cite{bsgNPmb} and
$1/m_c$ \cite{bsgNPmc} corrections.
From the formula above we obtain the theoretical result for 
BR($\Bsg$) in the SM which is given by
\beq
\BR^{\rm NLO}(B\to X_s\gamma)=(3.29\pm 0.33)\times 10^{-4}
\label{bsgSM}
\eeq
where the main theoretical uncertainty
comes from uncertainties in the SM input parameters,
namely $m_t,~\alpha_s(M_Z),
~\alpha_{em},~m_c/m_b,~m_b, V_{ij}$, and the small residual scale 
dependence.\footnote{
Recently in Ref.~\cite{neubert} the current method of
extracting the inclusive rate for \bsg from the currently published CLEO
data has been criticized arguing that the theoretical uncertainties have
so far been underestimated and only a precise measurement of the 
photon spectrum would be help in reducing these uncertainties.}
The central value in Eq.(\ref{bsgSM}) corresponds to the following 
central values for the SM parameters
$m_t^{\rm pole} \simeq m_t^{\rm \overline{MS}}(m_Z) \simeq 174\,\rm GeV$, 
$m_b^{\rm pole} = 4.8\,\rm GeV$, $m_c^{\rm pole} = 1.3\,\rm GeV$, 
$\mu_b = m_b$, $\alpha_s(m_Z) = 0.118$, $\alpha_e^{-1}(m_Z) = 128$, 
$\sin^2\theta_W = 0.23$ and a photon energy resolution corresponding 
to $\delta = 0.9$ is assumed. Note that in Eq.(\ref{BRSM})
the (small) $1/m_c$ corrections have not been included.

The SUSY contributions to the Wilson coefficients $C_{7,8}^{(0,1)}$ 
are obtained by calculating 
the \bsg and $b\to s g$ amplitudes at EW scale respectively.
The LO contributions to these amplitudes are given by the 1-loop 
magnetic-dipole and chromomagnetic dipole penguin diagrams respectively, 
mediated by charged Higgs boson, chargino, gluino, and neutralino exchanges. 
The corresponding results for these amplitudes can be found in Ref.\cite{BBMR}.
It is known that the charged Higgs contribution always interferes with
the SM contribution \cite{bsgSUSY,BBMR}. 
The chargino contribution could give rise to a 
substantial destructive interference with SM and charged Higgs amplitudes,
depending on the sign of $\mu$, the value of \tgb, and the mass difference
between the stop masses \cite{bsgSUSY,BBMR}.

We point out that 
the SUSY models with non--universal $A$--terms may induce non-negligible
contributions to the dipole operators $\tilde{Q}_{7,8}$
which have opposite chirality with respect to $Q_{7,8}$.
It is worth mentioning that 
these operators are also induced in the SM and in the MSSM with supergravity
scenario, but their contributions
are negligible being suppressed by terms of order $\O(m_s/m_b)$.
In particular in MSSM, due to the universality of the $A$--terms,
the gluino and chargino contributions 
to $\tilde{Q}_{7,8}$ turn out to be of order $\O(m_s/m_b)$.
This argument
does not hold in the models with non--universal $A$--terms and in
particular in our case.
It can be simply understood by using the mass insertion method \cite{GGMS}.
For instance, the gluino contributions to $Q_7$ and $\tilde{Q}_7$
operators are proportional to $(\delta_{LR}^d)_{23}\simeq
(S_{D_L}Y^{A\star}_d S_{D_R}^{\dag})_{23}/m_{\tilde{q}}^2$
and $(\delta_{RL}^d)_{23}\simeq 
(S_{D_R}Y^{A}_d S_{D_L}^{\dag})_{23}/m_{\tilde{q}}^2$ respectively.
Since the $A^D$ matrix is symmetric in model A and
$A^D_{ij}\simeq A^D_{ji}$ in model B, then 
$(\delta_{LR}^d)_{23}\simeq (\delta_{RL}^d)_{23}$.
Then in our case we should
consistently take into account the SUSY contributions to $\tilde{Q}_{7}$ in
\bsg.
Analogous considerations hold for the operator $\tilde{Q}_{8}$.

By taking into account the above considerations regarding the 
operators $\tilde{Q}_{7,8}$,
the new physics effects in \bsg
can be parametrized in a model independent way 
by introducing the so called $R_{7,8}$ and $\tilde{R}_{7,8}$ 
parameters defined at EW scale as
\beq
R_{7,8}=\frac{\left(C^{(0)}_{7,8}-C^{(0)SM}_{7,8}\right)}
{C_{7,8}^{(0)SM}},~~~
\tilde{R}_{7,8}=\frac{\tilde{C}_{7,8}^{(0)}}{C_{7,8}^{(0)SM}},
\label{R78}
\eeq 
where $C_{7,8}$ include the total contribution while $C_{7,8}^{SM}$ contains
only the SM ones. 
Note that in $\tilde{C}_{7,8}$, which are the corresponding Wilson coefficients
for $\tilde{Q}_{7,8}$ respectively,
 we have set to zero the SM contribution.
In Ref.~\cite{BBMR} only the expressions for the $R_{7,8}$ are given, for
completeness we report the corresponding expressions for $\tilde{R}_{7,8}$ 
in the appendix.

Inserting these definitions into the BR($\Bsg$) formula in Eq.(\ref{BRSM})
yields a general parametrization of the branching ratio in terms 
of the new physics contributions \cite{gabsarid} \footnote{Note that
the SM central value for BR($\Bsg$)
in Ref.~\cite{gabsarid} slightly differs from the result in Eq.(\ref{bsgSM}),
since in Ref.~\cite{gabsarid} the non-perturbative $\Lambda/m_c$ corrections 
have been included.}
\bea 
{\rm BR}(B\to X_s\gamma) &=& (3.29\pm 0.33)\times 10^{-4} 
\left(1 + 0.622 R_7 + 0.090(R_7^2+\tilde{R}_7^2)
\right.\nonumber \\ 
&+& \left.0.066 R_8 + 0.019 (R_7 R_8 + \tilde{R}_7 \tilde{R}_8)
+ 0.002(R_8^2+\tilde{R}_8^2)\right)~,
\label{bsgPAR}
\eea 
where the overall SM uncertainty has been factorized outside. 
We have checked explicitly that the result in Eq.(\ref{bsgPAR}) is
in agreement with the corresponding one used in Ref.~\cite{DTV}.
Recently 
the leading EW corrections in the SM have been included
in the $C_{7,8}$ coefficients \cite{MAR}. However,
we have not included them since they should
affect the results in a few percent which is within 
the theoretical uncertainties present in the SUSY sector.
In particular, in order to be consistent with the NLO calculations,
one should also include the corresponding two-loop QCD corrections to the 
SUSY amplitudes, namely $C_{7,8}^{(1)}$.
In this respect recent works \cite{bsgSUSYNLO}
have been done in this direction. 
They calculated the two-loop $\alpha_s$ corrections 
to the chargino and gluino amplitudes.
However these corrections can be consistently applied only in a
restricted SUSY scenario with low \tgb and large gluino masses,
since on the contrary other  
two-loop diagrams with no QCD interactions at all could become relevant.
In our work we are going to explore SUSY scenarios
in the full range of \tgb ($2\le\tan{\beta}\le m_t/m_b$) 
and gluino mass, so we 
do not include these corrections since 
they would not change our main conclusions.
%%%%%%%%%%%%%%%%%%%%%%%%%%%%%%%%%%%%%%%%%%%%%%%%%%%%%%%%%
\section{Numerical results and discussions}
In this section we present our results for the total branching ratio 
BR($\Bsg$) as a function of the fundamental
parameters of the soft-breaking sector of models A and B.
We start our discussion by analysing the constraints on 
these models set by the condition of vacuum stability 
and the experimental bounds on the SUSY particle spectrum.

We calculate the low energy SUSY spectrum by running the  
soft-breaking terms and the other SUSY parameters 
(by means of the renormalization group equations (RGE) of MSSM
generalized to the non--universal soft-breaking terms \cite{BBMR})
from the GUT scale $(\simeq 2\times 10^{16}$ GeV) 
to the EW scale ($\simeq M_Z$), by using
the boundary conditions given in section 2.
For fixed \tgb and sign of $\mu$, we restrict
the parameter space by imposing 
the present experimental bounds on the SUSY spectra \cite{PDG}.
Since the spectrum of these models is given in terms of few parameters
($m_{3/2},~\theta$ and/or $\Theta_i$)
we find that by requiring the lightest chargino
mass to be \mch$\ge ~90~{\rm GeV}$ implies that all the other SUSY particle
and the lightest Higgs masses are above their experimental bounds.

In order to avoid vacuum instabilities 
and color-charge breaking, we require that all 
the square scalar masses in Eqs.(\ref{scalar}), (\ref{scalarB})
should be positive.
All these constraints set strong restrictions on the parameter space 
of both models.
As pointed out in section 2,
in model A this leads to $\sin{\theta} > 1/\sqrt{2}\simeq 0.7$.
The bounds on $m_{3/2}$ from the lightest chargino mass
are as follows: for low \tgb (\tgb$=2$) $m_{3/2}\ge 80~(60)~{\rm GeV}$ 
for $\sin \theta \simeq 1/\sqrt{2}~(1)$. For large
\tgb these bounds are increased by $20$ GeV in both cases. 

In model B the non--universality is parameterized by the angle
$\theta$ and the $\Theta_i$'s.
The values $\Theta_i = 1/\sqrt{3}$ give
universal $A$ terms and scalar masses, the gaugino
masses are universal only at $\theta=\pi/6$. To avoid negative scalar masses
in this model one needs to impose the constraints
\begin{eqnarray}
\cos^2 \theta\ \Theta_i &<& 1/3,~(i=1,2,3)~, \no 
(1-\Theta_1^2) \cos^2 \theta &<& 2/3.
\label{thetaconstr}
\end{eqnarray}
In the limit of universal $A$ terms 
these two constraints are satisfied for any value of $\theta$.

Further constraints on the parameter space of model B are obtained
from the gaugino sector. As shown in Eq.(\ref{gauginoB}), for a non very large 
$m_{3/2}$, the limit of $\theta\simeq \pi/2$ can not be reached since 
$M_2$ would approach zero and the mass of the lightest chargino would be 
too small.  Moreover, the lower bound on the gluino mass and the condition of 
having the EW breaking at the correct scale, lead also to a lower bound on 
the angle $\theta$. In the case of universal $A$--terms we find that 
$\theta > 0.1$, while in the non--universal case Eqs.(\ref{thetaconstr}) 
set more severe constraints on $\theta$. For example, for 
$\Theta_1 \simeq 1$ we obtain that $0.9 \leq \theta < \pi/2$ and 
for $\theta$ close to $\pi/2$ the gravitino mass $m_{3/2}$ should be quite 
heavy (of order TeV) to make the lightest chargino higher than the
experimental bound. We find that these constraints together strongly reduce 
the allowed parameter space of model B.

We have checked that, in both models, 
the $B-\bar{B}$ mixing measurements do not set further constraints on the 
allowed ranges of the parameter space. This mixing, being a $\Delta B=2$ 
process, is proportional to $(\delta^d_{AB})_{13}\times
(\delta^d_{CD})_{13}$ where $A,B,C,D=(L,R)$. We have found that, in our case, 
the values of these mass insertions satisfy the constraints given in 
Ref.\cite{GGMS}.

\begin{figure}[tbc]
\centerline{
\begin{tabular}{c}
\epsfig{figure=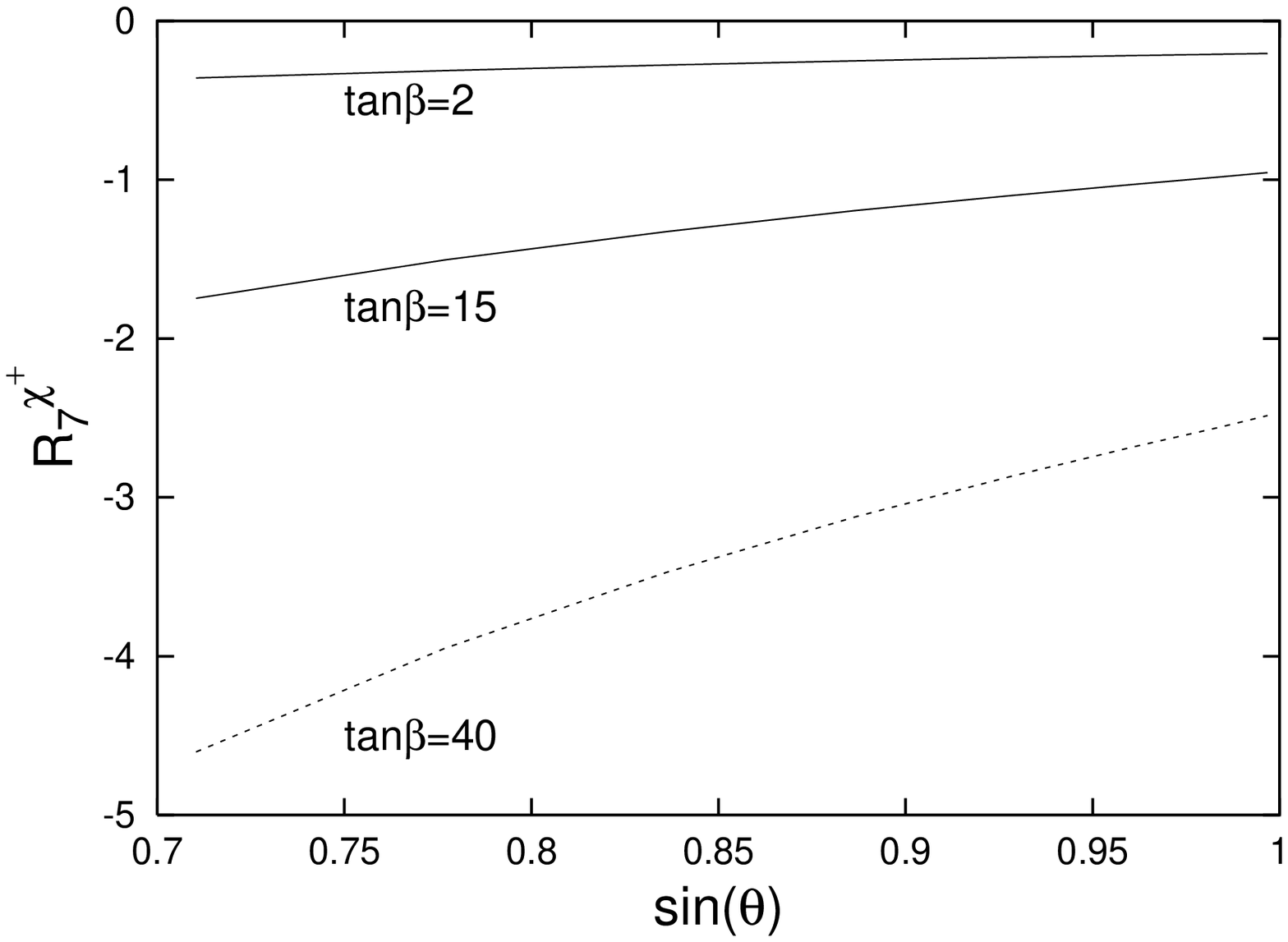,height=7cm,width=10.5cm,angle=0}\\
\epsfig{figure=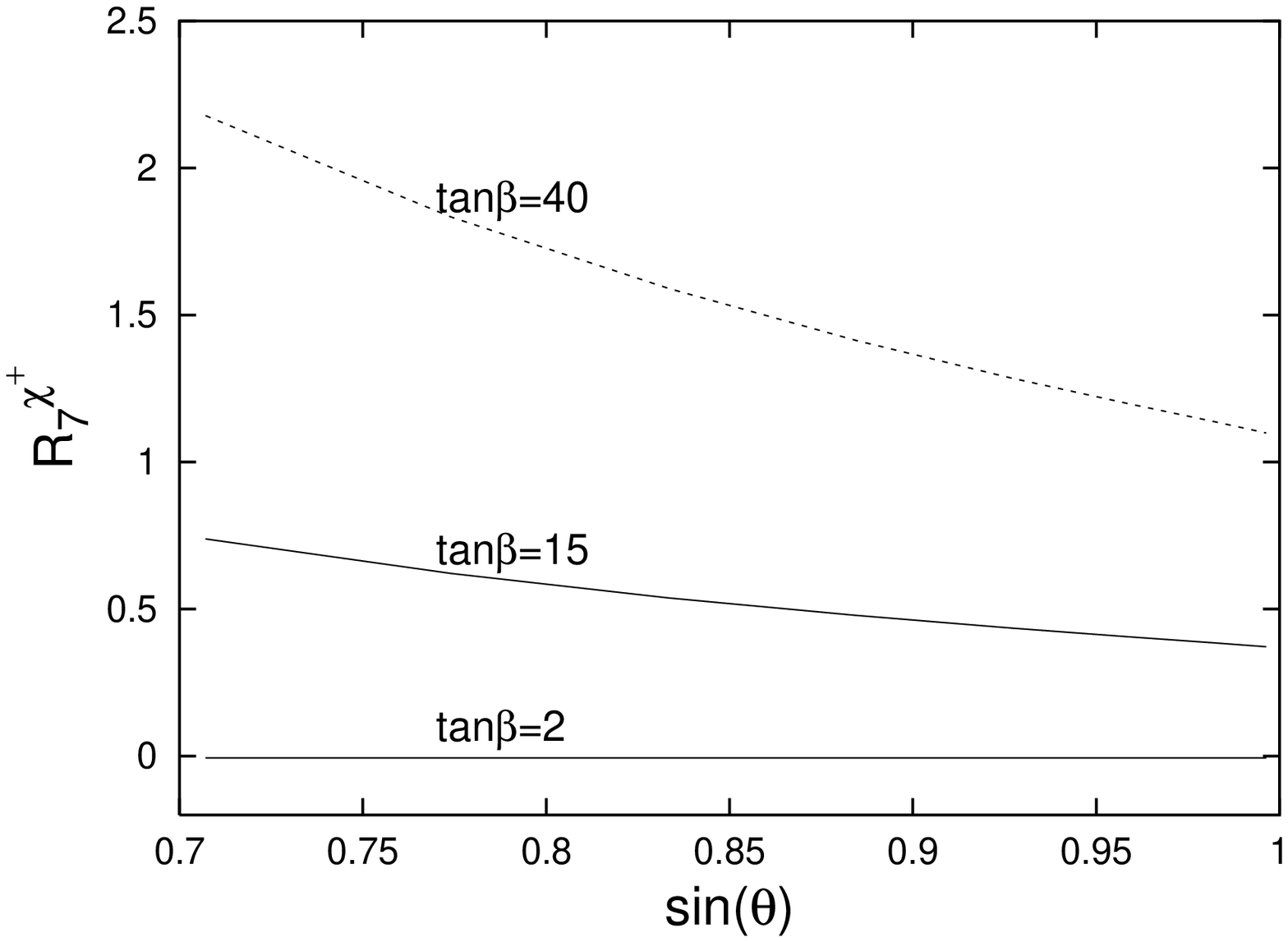,height=7cm,width=10.5cm,angle=0}
\end{tabular}
}
\caption{(Top) $R_7$ for chargino contribution
versus $\sin\theta$ in model A, for $\mu > 0$, $m_{3/2}=150$ GeV,
and for \tgb=2, 15, 40. (Bottom) 
The same as before, but for $\mu < 0$.}
\label{R7ch1}  
\label{R7ch2}  
\end{figure}

\begin{figure}[htbc]
\centerline{
\begin{tabular}{c}
\epsfig{figure=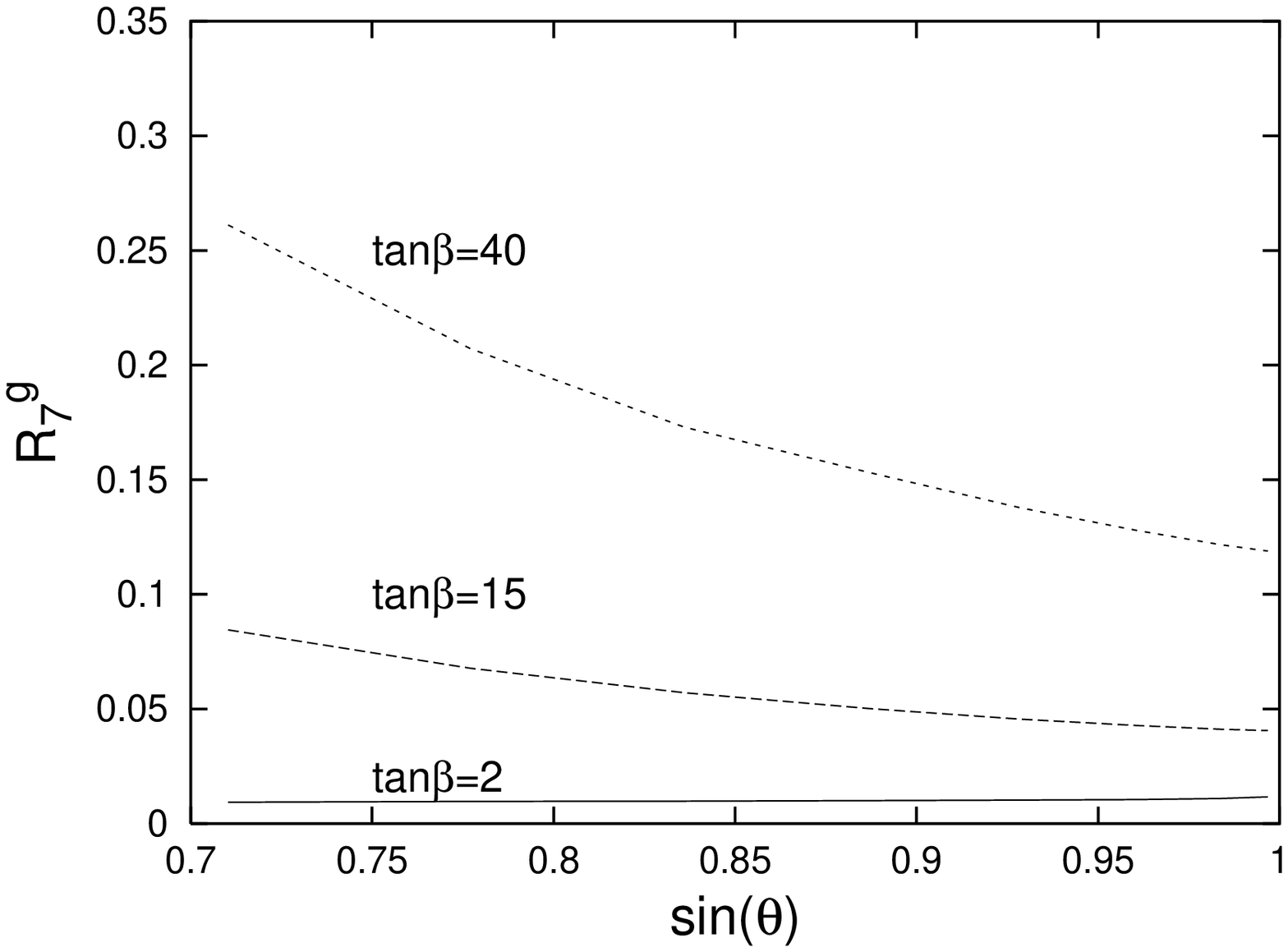,height=7cm,width=10.5cm,angle=0}\\
\epsfig{figure=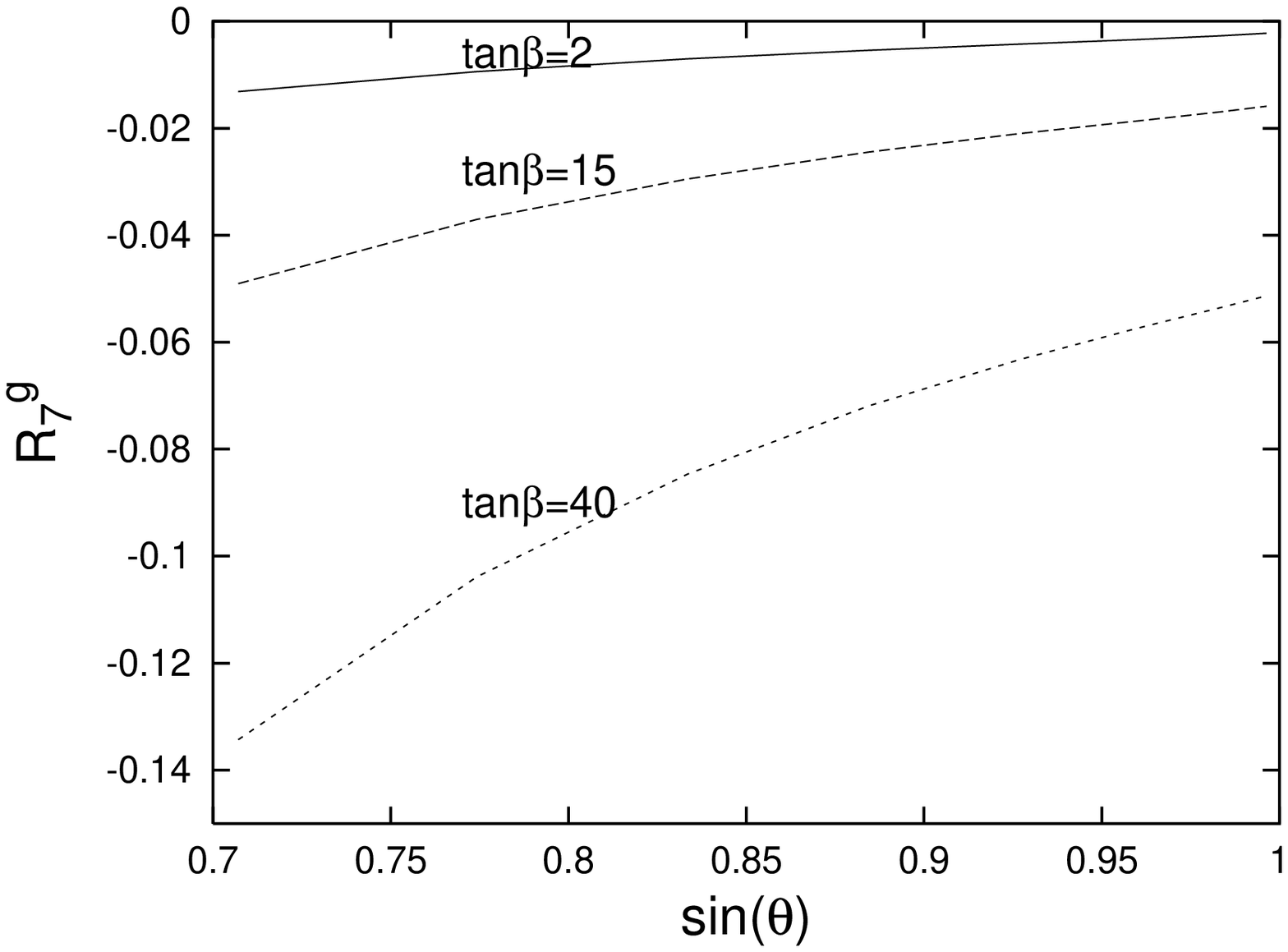,height=7cm,width=10.5cm,angle=0}
\end{tabular}
}
\caption{
(Top) $R_7$ for the gluino contribution 
versus $\sin\theta$ in model A, for $\mu > 0$, 
$m_{3/2}=150$ GeV, and for \tgb=2, 15, 40. (Bottom) 
The same as before, but for $\mu < 0$.}
\label{R7gl1}  
\label{R7gl2}  
\end{figure}

\begin{figure}[htbc]
\centerline{
\epsfig{figure=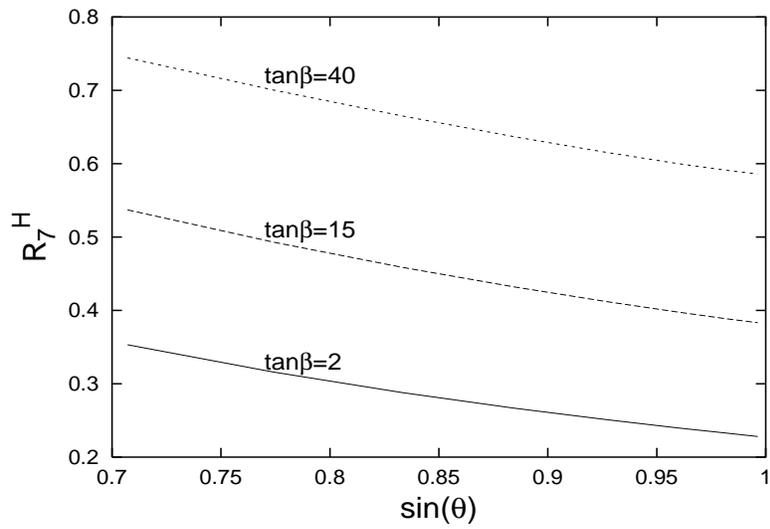,height=7cm,width=10.5cm,angle=0}}
\caption{ $R_7$ for charged Higgs contribution
versus $\sin\theta$ in model A, $m_{3/2}=150$ GeV,
and for \tgb=2, 15, 40.}
\label{R7h1}  
\end{figure}

\begin{figure}[htcb]
\centerline{
\begin{tabular}{c}
\epsfig{figure=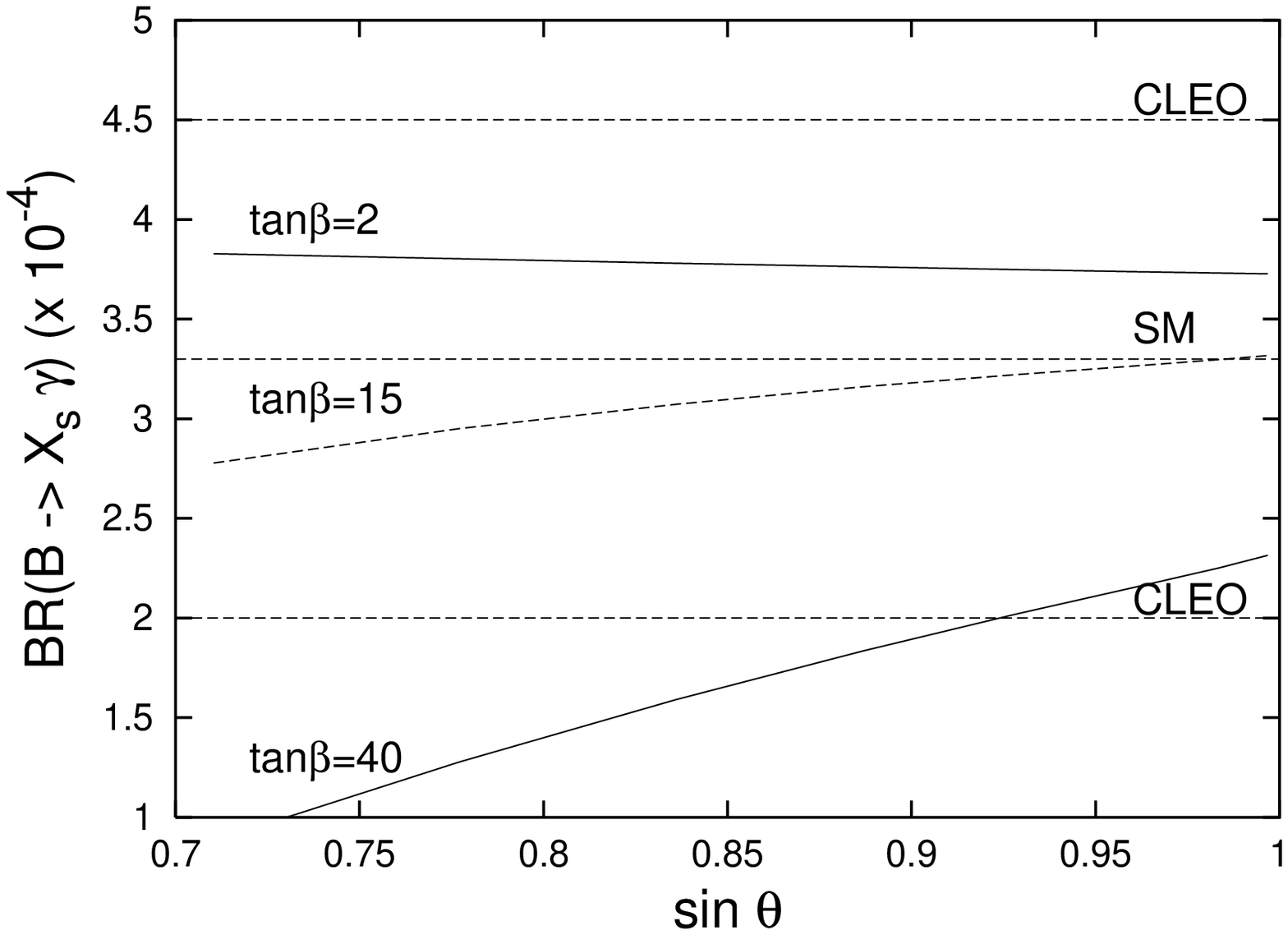,height=7cm,width=10.5cm}\\
\epsfig{figure=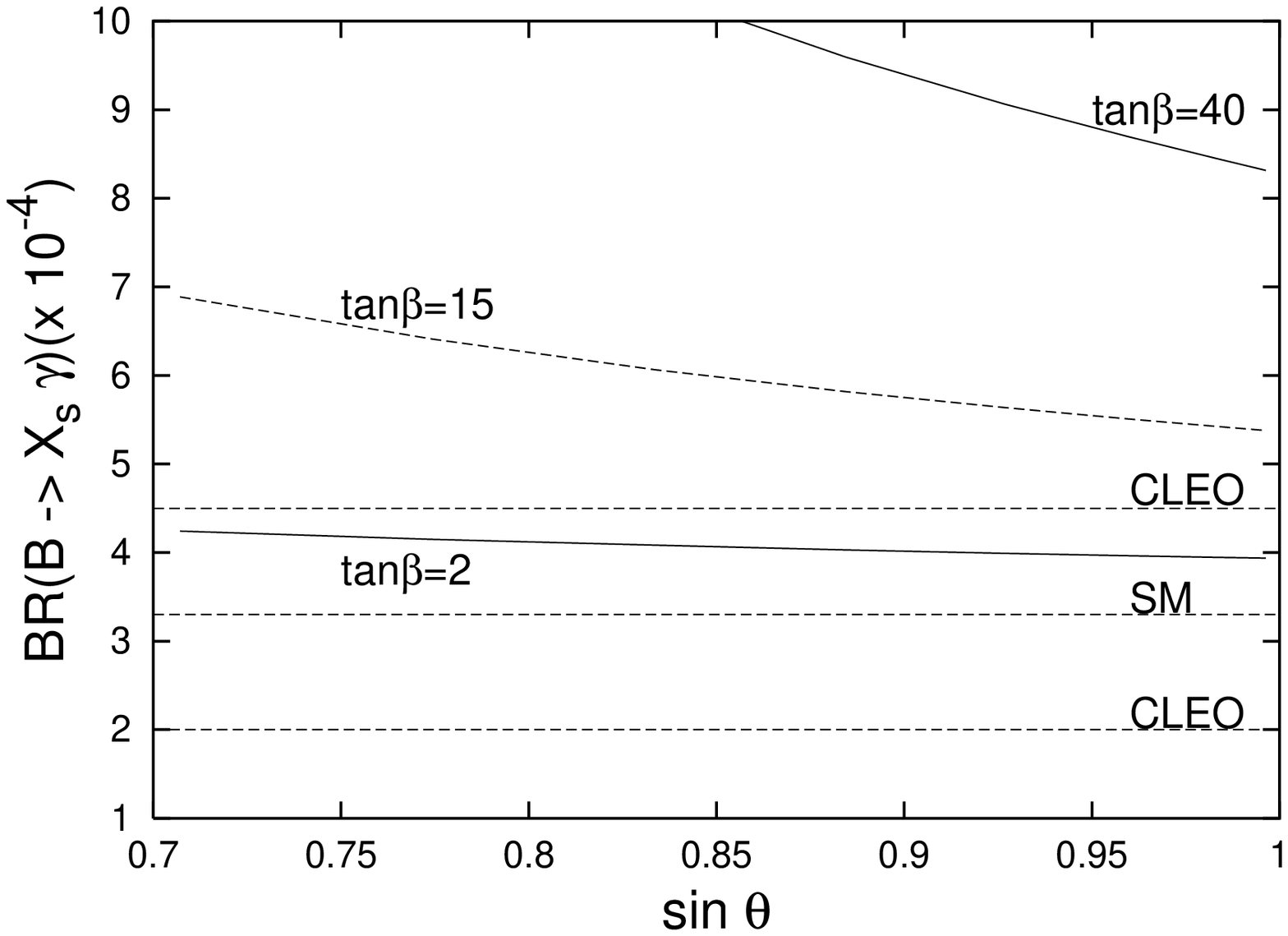,height=7cm,width=10.5cm}
\end{tabular}
}
\caption{(Top) The BR($\Bsg$) versus $\sin\theta$ in model A, 
for $\mu>0$, $m_{3/2}=150$ GeV and \tgb=2, 15, 40.
(Bottom) The same as before, but for $\mu<0$.}
\label{BRA1}  
\label{BRA2}  
\end{figure}

\begin{figure}[htbc]
\centerline{
\begin{tabular}{c}
\epsfig{figure=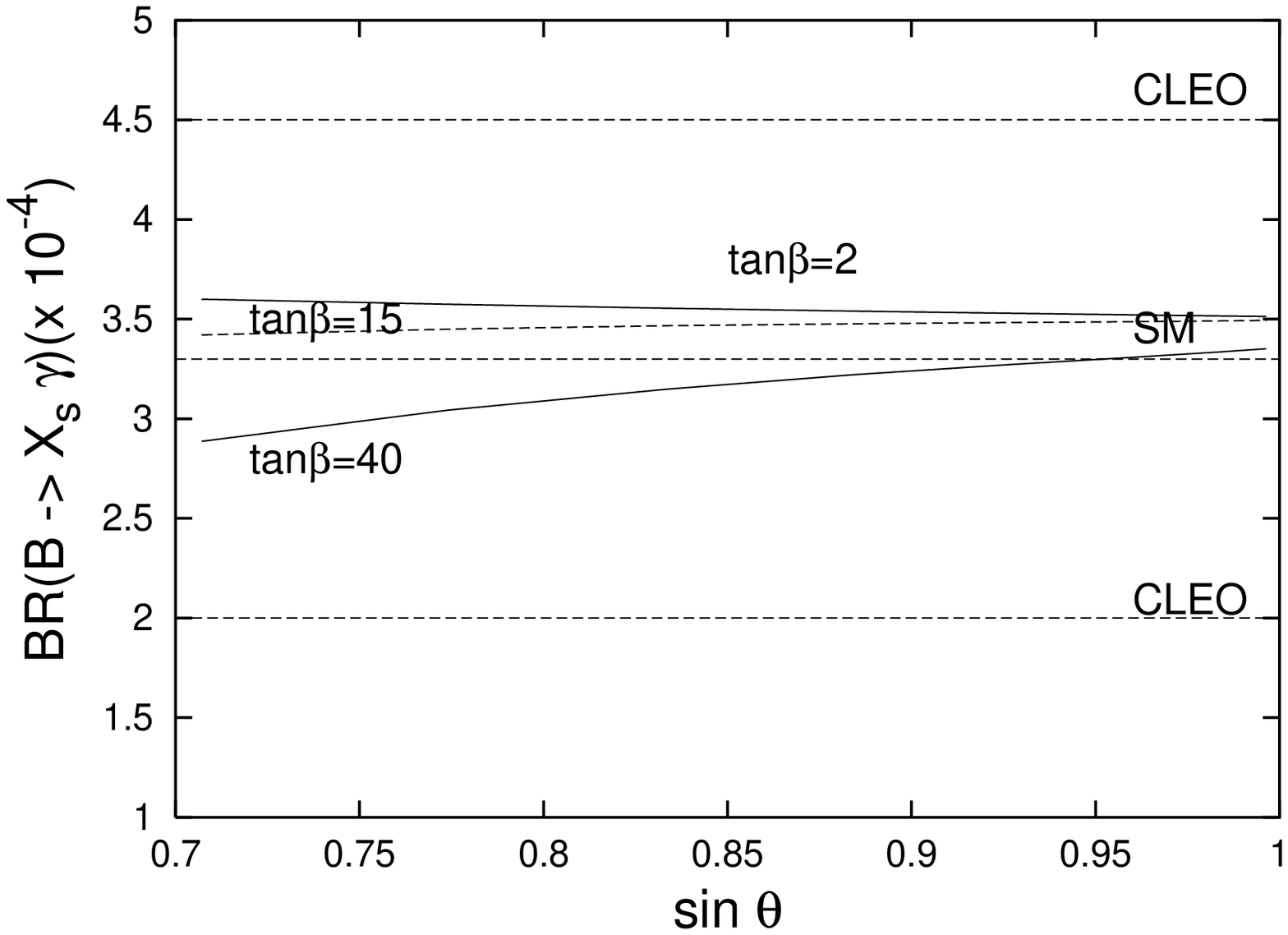,height=7cm,width=10.5cm,angle=0}\\
\epsfig{figure=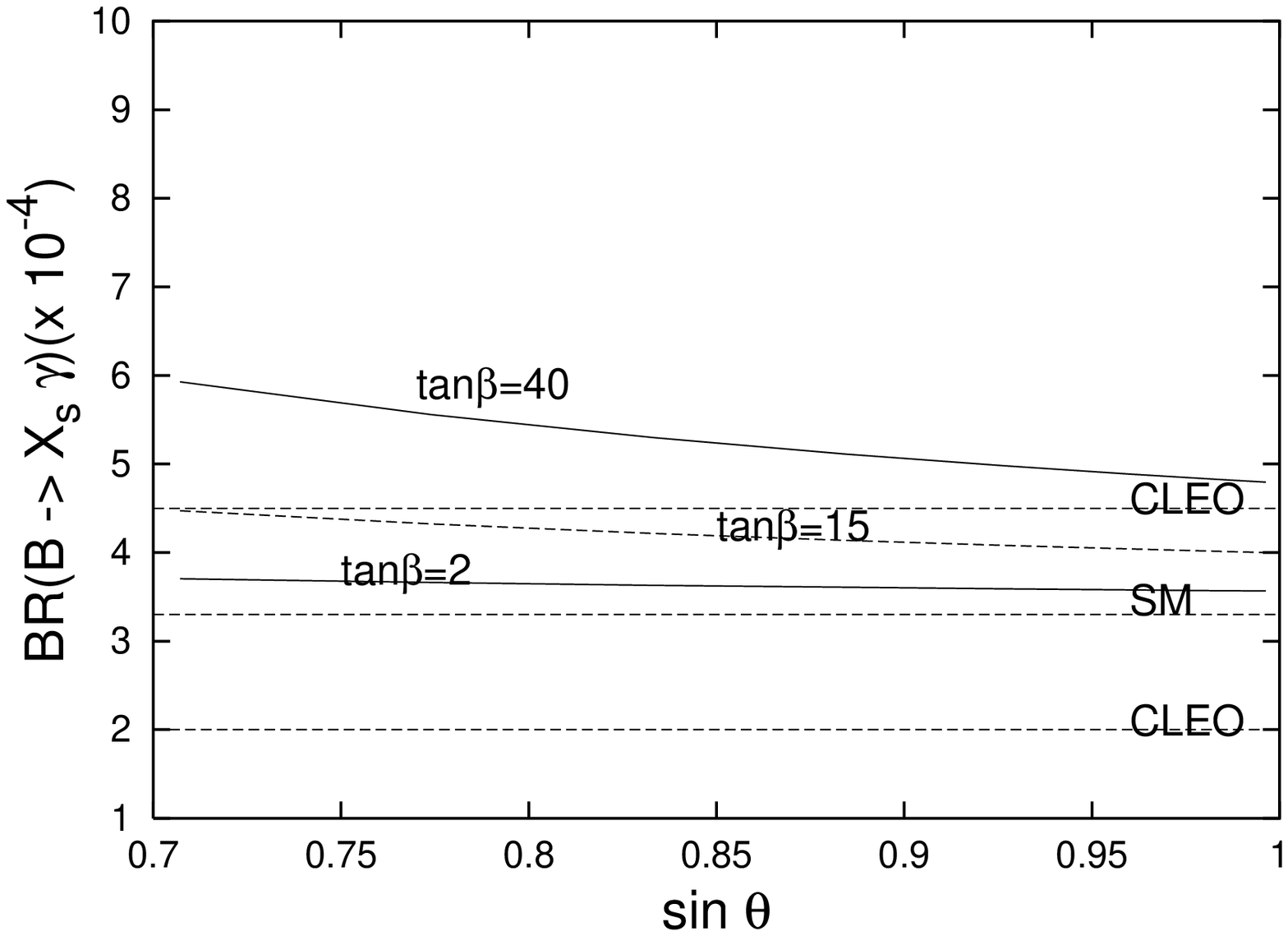,height=7cm,width=10.5cm,angle=0}
\end{tabular}
}
\caption{ 
(Top) The BR($\Bsg$) versus $\sin\theta$ in model A, for
$\mu > 0$, $m_{3/2}=300$ GeV and \tgb=2, 15, 40.
(Bottom) The same as before, but for $\mu < 0$.}
\label{BRA3}  
\label{BRA4}  
\end{figure}

\begin{figure}[tbc]
\centerline{
\begin{tabular}{c}
\epsfig{figure=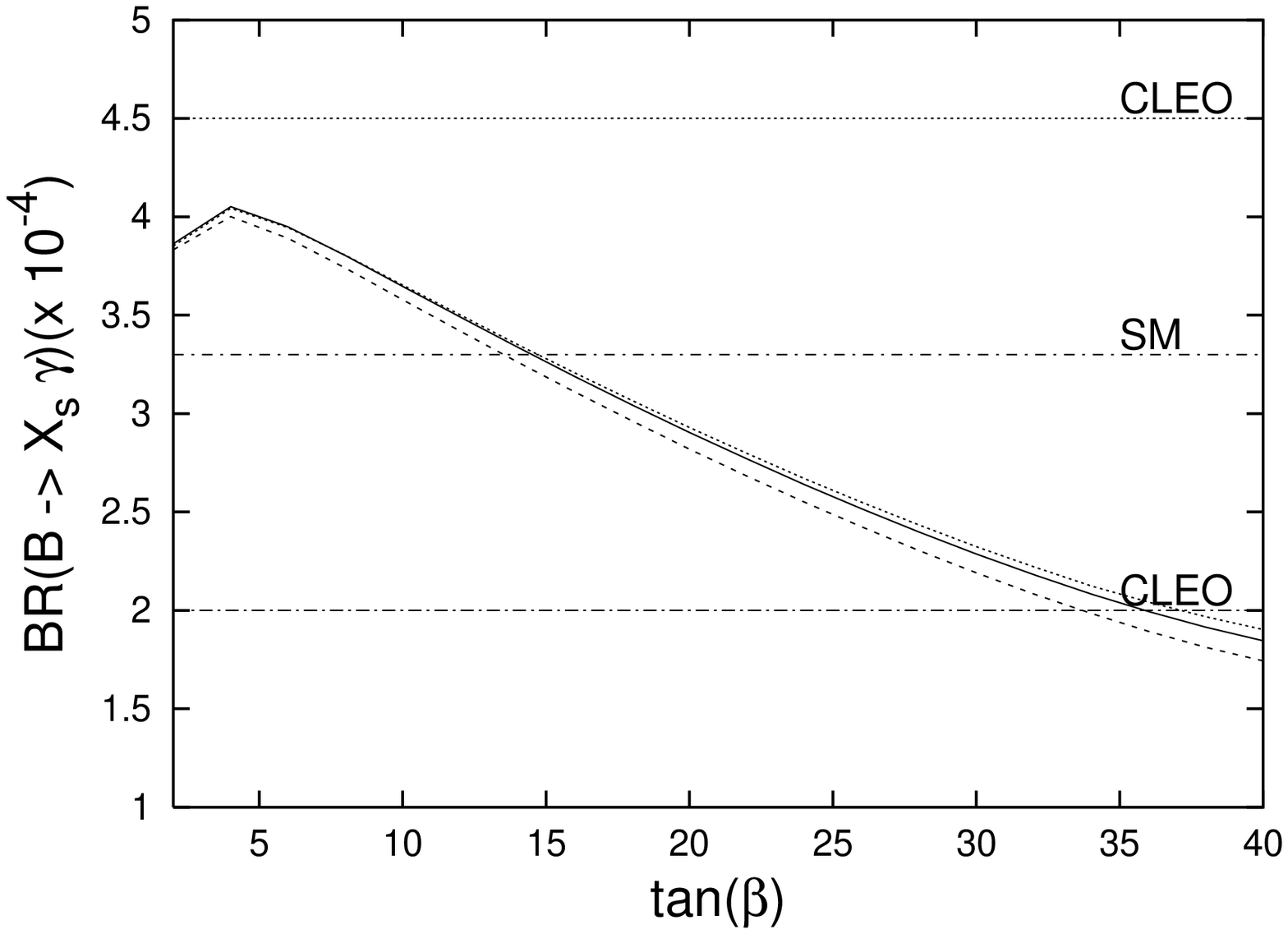,height=7cm,width=10.5cm,angle=0}\\
\epsfig{figure=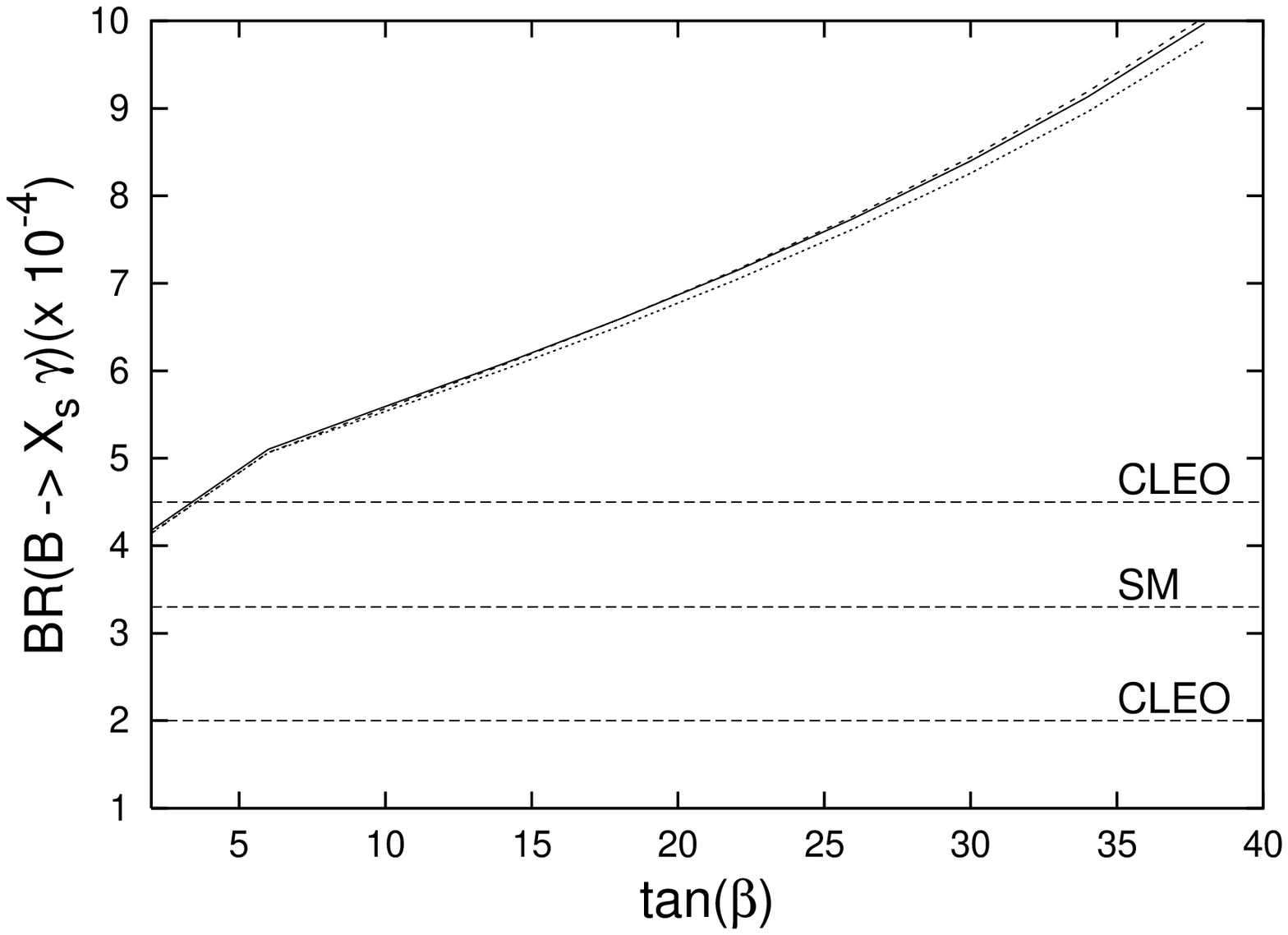,height=7cm,width=10.5cm,angle=0}
\end{tabular}
}
\caption{(Top) The branching ratio BR($\Bsg$) versus 
$\tan \beta$ in model B, for $\mu > 0$, $m_{3/2}=150$ GeV,
and for some values of $(\Theta_1,\Theta_2)=
(1/\sqrt{3},1/\sqrt{3}),~(0.9,0.2),~(0.6,0.2)$, corresponding to the
continuos, dashed, and dot-dashed lines respectively.
(Bottom) The same as before, but for $\mu < 0$.}
\label{BRB1}  
\label{BRB2}  
\end{figure}

Our results for the partial SUSY amplitude contributions 
are presented in Figs.[\ref{R7ch1}-\ref{R7h1}]. The total branching 
ratio BR($\Bsg$) is shown in Figs.[\ref{BRA1}-\ref{BRA4}] 
and [\ref{BRB1}] for model A and B respectively.
In Figs.[\ref{R7ch1}-\ref{R7h1}] we show, for model A, 
the individual SUSY contributions to the $R_{7}$ variable, (see  Eq.(\ref{R78})
for its definition) versus $\sin{\theta}$. 
We see that the chargino and charged 
Higgs contributions give the dominant effect in all the range of $\theta$, 
while the gluino is sub--dominant, such as in the universal case. 
For model B, as expected from its tightly constrained parameter space, 
we have found that the results of the separate amplitude 
contributions do not differ from the corresponding ones in the universal 
scenario, and will not be presented here.  

We can understand this behavior by using the mass insertion method.
The gluino amplitude gets two leading contributions: one
is proportional to the single--mass--insertion $(\delta_{LR}^d)_{23}$
and the another one to the double--mass--insertion, namely
$(\delta_{LR}^d)_{22}(\delta_{LL}^d)_{23}$. 
In the low and intermediate \tgb regions these two mass insertions
are comparable, so a possible destructive or constructive 
interference between them may appear depending on the sign of $\mu$.
In the large \tgb region
the double mass insertion becomes dominant, since 
$(\delta_{LR}^d)_{22}$ is proportional to $\mu\tan{\beta}$ and
the amplitude (normalized to the SM one) is
\beq
\Frac{A_{\tilde{g}}}{A_{SM}} \propto \Frac{\alpha_s}{\alpha_W V_{32}}
\tan{\beta}~(m_W^2~ M_{\tilde{g}}~\mu)~ \Frac{M^2_{\tilde{b}_L\tilde{s}_L}}
{m_{\tilde{q}}^6},
\eeq
where $m_{\tilde{q}}$ is the average squark mass in the down sector
and $M^2_{\tilde{b}_L\tilde{s}_L}$ is the off diagonal element of 
the down-squark mass matrix defined in Eqs.(\ref{sqmass},\ref{sqmass2}).
In this way, the sensitivity of the gluino amplitude to 
the sign of $\mu$ for large \tgb can be understood. 
The chargino amplitude, like in the gluino case, is enhanced by \tgb and
the dominant contributions to it are  
the (Higgsino) $A_{\tilde{h}^-}$ and (Wino-Higgsino) 
$A_{\tilde{W}\tilde{h}^-}$ amplitudes, given by
\bea
\Frac{A_{\tilde{h}^-}}{A_{SM}}&\propto& 
\tan{\beta}~(\mu~ m_t) \Frac{M^2_{\tilde{t}_L\tilde{t}_R}}
{m_{\tilde{t}_L}^2m_{\tilde{t}_R}^2},
\\
\Frac{A_{\tilde{W}{\tilde{h}^-}}}{A_{SM}}&\propto& \Frac{1}{V_{32}}
\tan{\beta}~(\mu~ M_2)~ \Frac{M^2_{\tilde{t}_L\tilde{t}_L}}
{m_{\tilde{t}_L}^2m_{\tilde{t}_R}^2}.
\eea
We see that these contributions are large with respect to 
the gluino one, such as in the universal case of MSSM,
mainly because the mass insertions get a
large enhancement of the light-stop mass $m_{\tilde{t}_L}^2$
in the denominator, unlike in the gluino case where the 
$1/m_{\tilde{q}}^6$ suppression is effective.

Regarding the contribution of the $\tilde{Q}_{7,8}$ operators to the
total branching ratio, we checked that their effect is negligible, 
almost an order of magnitude smaller than the total contribution to
$Q_{7,8}$. This  can be explained by observing that the dominant 
effect to these operators comes from the gluino amplitude
which is much smaller than the chargino or charged Higgs one.

In Figs. [\ref{BRA1},\ref{BRA4}] we plot the results
for the branching ratio BR($\Bsg$), in model A,  versus $\sin{\theta}$  
for different values of \tgb, the sign of $\mu$ 
and for two representative values of gravitino mass \mgrav, 
namely $m_{3/2}=150,~300$ GeV.
The main message arising from these results 
is that the sensitivity of BR($\Bsg$) respect to $\sin \theta$
increases with \tgb. 
In particular for the low \tgb region the \bsg result
does not differ significantly from the universal case. In the large
\tgb region, $\tgb=15-40$, the CLEO measurement of \bsg set severe 
constraints on the angle  $\theta$ for low gravitino masses.
For $\mu<0$ almost the whole range of parameter
space is excluded as shown in Fig. [\ref{BRA2},\ref{BRA4}].

Comparing Top and Bottom plots in  Figs.[\ref{BRA1},\ref{BRA3}], 
we see that that for positive (negative) 
sign of $\mu$ the branching ratio BR($\Bsg$) decreases (increase) when
the departure from the universality increases ($\sin \theta \to 1/\sqrt{2}$). 
Clearly, due to decoupling effects, the deviations from universality
tend to be reduced for large gravitino masses, as can be seen by
comparing the plots at $m_{3/2}=150~{\rm GeV}$ with the corresponding
ones at $m_{3/2}=300~{\rm GeV}$.

Now we discuss results in model B. In Fig.[\ref{BRB1}] we 
plot the branching ratio BR($\Bsg$) versus \tgb for three different 
values of $\Theta_1,\Theta_2$ (see  the figure caption)
 which are representative examples for universal and highly non--universal cases.
From these figures it is clear that BR($\Bsg$) is not very sensitive to the
values of $\Theta_i$'s parameters, even at very large \tgb, unlike model A.
The constraints from CLEO measurement are almost the same in the
universal and non--universal cases. For $\mu > 0$
the branching ratio is constrained from the lower bound of CLEO only
at very large \tgb, while for $\mu < 0$ the branching ratio is 
almost excluded except at low \tgb.

%%%%%%%%%%%%%%%%%%%%%%%%%%%%%%%%%%%%%%%%%%%%%%%%%%%%%%%%%%%%%%%%%%%%%%%%%%%%
\section{Conclusions}
\label{sec:conclusions}
The recent CP violation measurements of 
$\varepsilon^{\prime}/\varepsilon$ indicate a large deviation from the
SM predictions which might be interpreted like a signal of new 
CP violation sources beyond the SM. 
It is unlikely that the minimal supersymmetric standard model
with universal boundary conditions at GUT scale can explain these
large enhancements in the direct CP violations. On the contrary, 
non-minimal SUSY scenarios with non--universal $A$--terms, derived from
some string inspired models, have been found effective 
in explaining large values for  $\varepsilon^{\prime}/\varepsilon$ 
while keeping the electric dipole moments below the experimental bounds.

In this paper we have considered two models based on weakly coupled 
heterotic string (model A) and type I string theories (model B).
In this framework we carefully analysed the constraints set by the \bsg decay
on these two models by taking into account the relevant set of SUSY diagrams.
In the calculation of the total branching ratio we take into account
the NLO QCD corrections to the SM.
We found that in the model based on the weakly coupled 
heterotic string, the \bsg branching ratio is more sensitive 
to the non--universality at large \tgb and that the dominant SUSY contribution
comes from the chargino amplitude for any value of \tgb.
For type I string-derived model,
we found that the sensitivity of the \bsg branching ratio 
to the non--universality parameters $\theta$ and $\Theta_i$ 
is quite weak. The main reason for this weakness
is because in this model 
the allowed ranges for these parameters are strongly constrained 
by the vacuum stability bounds and
the experimental limits on the lightest chargino mass.

We conclude that the recent CLEO measurements on the total inclusive
B meson branching ratio BR($\Bsg$) do not set severe constraints on the
non--universality of these models. 
Moreover the constraints set on \tgb and gravitino mass
are almost the same as in the universal case.
In this respect we have found that the parameter regions 
which are important for generating sizeable contributions to
$\varepsilon^{\prime}/\varepsilon$ \cite{khalil1,khalil2,vives}, 
in particular the low
\tgb regions, are not excluded by \bsg decay.
%%%%%%%%%%%%%%%%%%%%%%%%%%%%%%%%%%%%%%%%%%%%%
%%%%%%%%%%%%%%%%%%%%%%%%%%%%%%%%%%%%%%%%%%%%%%%%%%%%%%%%%%%%%%%%%%%%%%%%
\section*{Acknowledgments}
We would like to thank C. Mu\~{n}oz for useful discussions.
S. K. acknowledges the financial support of a Spanish Ministerio de 
Educacion y Cultura research grant.
E.G. acknowledges the financial support of the TMR network, project
``Physics beyond the standard model'', FMRX-CT96-0090.
The work of E.T. was supported by DGICYT grant AEN97-1678.
%%%%%%%%%%%%%%%%%%%%%%%%%%%%%%%%%%%%%%%%%%%%%%%%%%%%%%%%%%%%%%
\section*{Appendix}
Here we give the expressions for the dominant SUSY contributions
to $\tilde{R}_{7,8}$ defined in Eq.(\ref{R78}), namely the chargino and gluino ones,
in the approximation $\O(m_s/m_b)=0$
\bea
\tilde{R}_{7,8}&=&\tilde{R}_{7,8}^{\chi}+\tilde{R}_{7,8}^{\tilde{g}}\no
\tilde{R}^{\chi}_7&=&-
\frac{2}{3 V_{32}^{\star}V_{33}x_{tW}F_7(x_{tW})}\no
&\times&\sum_{I=1}^{2}\sum_{k=1}^6x_{W\tilde{u}_k}
\left(X_I^{L}\right)_{k3}
\left(X_I^{R}\right)^{\star}_{k 2}\frac{m_{\chi_I}}{m_b}\left(
F_3(x_{\chi_I\tilde{u}_k})+\frac{2}{3}F_4(x_{\chi_I\tilde{u}_k})\right)
\no
%------------
\tilde{R}^{\chi}_8 &=&-
\frac{2}{3 V_{32}^{\star}V_{33}x_{tW}F_1(x_{tW})}
\sum_{I=1}^{2}\sum_{k=1}^6x_{W\tilde{u}_k}
\left(X_I^{L}\right)_{k3}
\left(X_I^{R}\right)^{\star}_{k 2} \frac{m_{\chi_I}}{m_b}
F_4(x_{\chi_I\tilde{u}_k})\no
%------------------------------
\tilde{R}^{\tilde{g}}_7&=&-
\frac{16\alpha_S}{27 \alpha_W V_{32}^{\star}V_{33}x_{tW}F_7(x_{tW})}
\sum_{k=1}^6 x_{W\tilde{d}_k}
\left(\Gamma^{D_L}\right)_{k3}
\left(\Gamma^{D_R}\right)^{\star}_{k 2}\frac{m_{\tilde{g}}}{m_b}
F_4(x_{\tilde{g}\tilde{d}_k})\no
%------------
\tilde{R}^{\tilde{g}}_8&=&-
\frac{2\alpha_S}{9 \alpha_W V_{32}^{\star}V_{33}x_{tW}F_1(x_{tW})}\no
&\times&\sum_{k=1}^6 x_{W\tilde{d}_k}
\left(\Gamma^{D_L}\right)_{k3}
\left(\Gamma^{D_R}\right)^{\star}_{k 2}\frac{m_{\tilde{g}}}{m_b}
\left(9F_3(x_{\tilde{g}\tilde{d}_k})+F_4(x_{\tilde{g}\tilde{d}_k})
\right),
\eea
with
\bea
\left(X_I^L\right)_{k i}&=&-{\bf V}_{I 1}\left(\Gamma^{U_L}~ V\right)_{ki} +
\frac{1}{\sqrt{2}m_W\sin{\beta}}
{\bf V}_{I 2}\left(\Gamma^{U_R}~ M_U~ V\right)_{k i}\no
\left(X_I^R\right)_{ki}&=&\frac{1}{\sqrt{2}m_W\cos{\beta}}
{\bf U}_{I 2}\left(\Gamma^{U_L}~V~ M_D\right)_{ki}
\eea
where $x_{ij}\equiv m_i^2/m_j^2$,
$F_7(x)=\Frac{2}{3} F_1(x)+F_2(x)$, ${\bf U}$ and 
${\bf V}$ are the $2\times 2$ 
diagonalization matrices for the chargino mass matrix defined as in 
Ref.~\cite{BBMR}, and the $6\times 6$ matrices
$\Gamma^{(U,D)}\equiv 
\left[\Gamma^{(D_L,U_L)}_{6\times 3}~,~\Gamma^{(D_R,U_R)}_{6\times 3}\right]$
respectively diagonalize the Up- and Down-squark mass matrices in 
Eqs.(\ref{sqmass},\ref{sqmass2}).
The expressions for the functions $F_{i}(x)$ can be found in Ref.\cite{BBMR}.

%%%%%%%%%%%%%%%%%%%%%%%%%%%%%%%%%%%%%%%%%%%%%%%%%%%%%%%%%%%%%%%%%

%%%%%%%%%%%%%%%%%%%%%%%%%%%%%%%%%%%%%%%%%%%%%%%%%%%%%%%%%%%%%%%%%%%%%%%%%
\newpage

\newpage

\begin{thebibliography}{99}
\bibitem{nonuniv}
 M. Brhlik, L. Everett, G. L. Kane, S. F. King, and O. Lebedev,
\prl{84}{2000}{3041};
R.~Barbieri, R.~Contino, and A.~Strumia, hep-ph/9908255;
K.~Babu, B.~Dutta, and R.N.~Mohapatra, \prd{61}{2000}{091701}.
%-----------------------
\bibitem{abel}
S.~Abel and J.~Frere, \prd{55}{1997}{1623}.
%-----------------------
\bibitem{khalil1}
S.~Khalil, T.~Kobayashi, and A.~Masiero, \prd{60}{1999}{075003}.
%-----------------------
\bibitem{khalil2}
S.~Khalil and T.~Kobayashi, \plb{460}{1999}{341}.
%-----------------------
\bibitem{vives}
S. Khalil, T. Kobayashi, and O. Vives,  hep-ph/0003086.
%----------------------
\bibitem{ibanezlust}
L. E. \Ibanez and D. Lust, \npb{382}{1992}{305}.
%----------------------
\bibitem{kapl}
V. S. Kaplunovsky and J. Louis, \plb{306}{1993}{269}.
%-----------------------
\bibitem{ibanez1}
A. Brignole, L. E. \Ibanez, and C. \Munoz, \npb{422}{1994}{125},
Erratum-ibid. {\bf B 436}~(1995) 747.
%-----------------------
\bibitem{ibanez2}
L. E. \Ibanez, C. \Munoz, and S. Rigolin, \npb{553}{1999}{43}.
%-----------------------
\bibitem{CP1}
A. Alavi--Harati {\it et al.} (KTeV Coll.), \prl{83}{1999}{22}.
%-----------------------
\bibitem{CP2}
V. Fanti {\it et al.} (NA48 Coll.), \plb{465}{1999}{335}.
%----------------------
\bibitem{CP3}
G. D'Agostini, hep-ex/9910036.
%-----------------------
\bibitem{epsp1}
A. Buras, M. Jamin, and M.E: Lautenbacher, \npb{408}{1993}{209};
M. Ciuchini, E. Franco, G. Martinelli, and L. Reina, \npb{415}{1994}{403};
S. Bosh, A.J. Buras, M. Gorbahn, S. Jager, M. Jamin, 
M.E. Lautenbacher, and L. Silvestrini, hep-ph/9904408;
M. Ciuchini, E. Franco, L. Giusti, V. Lubicz, and 
G. Martinelli, hep-ph/9910237; M. Jamin, hep-ph/9911390.
%-----------------------
\bibitem{epsp2}
S. Bertolini, M. Fabbrichesi, and J.O. Eeg, hep-ph/9802405;
T. Hambye, G.O. Kohler, E.A. Paschos, and P.H. Soldan, 
hep-ph/9906434; J.Bijnens, and J.Prades, JHEP 01, (1999) 023;
E. Pallante and A. Pich, hep-ph/9911233.
%-----------------------
\bibitem{GG}
E. Gabrielli and G.F. Giudice, \npb{433}{1995}{3};
Erratum-ibid. {\bf B~507} (1997) 549.
%-----------------------
\bibitem{masieromur} 
A. Masiero and H. Murayama, \prl{83}{1999}{907}.
%-----------------------
\bibitem{demir}
D.A.~Demir, A.~Masiero, and O.~Vives, hep-ph/9911337;
D.~A.~Demir, A.~Masiero, and O.~Vives, \prd{61}{2000}{075009}.
%-----------------------
\bibitem{barr}
S.~Barr and S.~Khalil, \prd{61}{2000}{035005}.
%------------------------
\bibitem{GGMS}
F. Gabbiani, E. Gabrielli, A. Masiero, and L. Silvestrini,
\npb{477}{1996}{321}.

%-----------------------
\bibitem{bsgSUSY} 
W. S. Hou and R.S. Willey \plb{202}{1988}{591};
T. G. Rizzo \prd{38}{1988}{820};
V. Barger, M.S. Berger, and R.J.N. Phillips, \prl{70}{1993}{1368};
J. L. Hewett, \prl{70}{1993}{1045};
R. Barbieri and G.F. Giudice, \plb{309}{1993}{86};
J. L. Lopez, D. V. Nanopoulos, and G. T. Park, \prd{48}{1993}{974};
Y. Okada, \plb{315}{1993}{119};
R. Garisto and J.N. Ng, \plb{315}{1993}{372}; 
M.A. Diaz, \plb{322}{1994}{207}; F.M. Borzumati, Z. Phys. 
{\bf C 63} (1994) 291;
P. Nath and R. Arnowitt, \plb{336}{1994}{395};
S. Bertolini and F. Vissani, Z. Phys. {\bf C 67} (1995) 513;
N.G. Deshpande, B. Dutta, and S. Oh, \prd{56}{1997}{519};
S. Khalil, A. Masiero, and Q. Shafi, \prd{56}{1997}{5754};
T. Blazek and S. Raby,\prd{59}{1999}{095002}.

%------------------------%----------------------
\bibitem{BBMR}
S. Bertolini, F. Borzumati, A. Masiero, and G. Ridolfi,
\npb{353}{1991}{591}.











%------------------------
\bibitem{type1}
L. Everett, G.L. Kane, and S.F. King, hep-ph/0005204;
E. Accomando, R. Arnowitt, and B. Dutta, \prd{61}{2000}{075010};
M. Brhlik, L. Everett, G.L. Kane, and J. Lykken, hep-ph/9908326;
L. E. \Ibanez and F. Quevedo, {\bf JHEP} (1999) 9910:001;
M. Brhlik, L. Everett, G.L. Kane, and J. Lykken (Fermilab)
\prl{83}{1999}{2124};
A. Corsetti and P. Nath, hep-ph/0003186;
S. Khalil, hep-ph/9910408; 
T. Ibrahim and P. Nath, \prd{61}{2000}{093004}.
%-----------------------
\bibitem{misiak1}
P. Cho, M. Misiak, and D. Wyler, \prd{54}{1996}{3329}.
%------------------
\bibitem{CLEO}
S. Ahmed {\it et al.}, (CLEO Collaboration), CLEO-CONF-99-10, 
hep-ex/9908022.
%------------------
\bibitem{ALEPH}
R. Barate {\it et al.} (ALEPH Collaboration), \plb{429}{1998}{169}.
%-----------------------
\bibitem{bsgNLO}
K. Chetyrkin, M. Misiak, and M. Munz, Phys. Lett. B {\bf 400}, 206 (1997);
A. J. Buras, A. Kwiatkowski, and N. Pott, Phys. Lett. B {\bf 414}, 157 (1997);
C. Greub and T. Hurth, \prd{54}{1996}{3350}; \prd{56}{1997}{2934}.




%-----------------------
\bibitem{PDG}
Review of Particle Physics, Eur. Phys. J. {\bf C3} (1998) 1.
%-----------------------
\bibitem{bsgNPmb} A.F. Falk, M. Luke and M. Savage, \prd{49}{1994}{3367}.
%------------------------
\bibitem{bsgNPmc} M.B. Voloshin, \plb{\bf 397}{1997}{275};
A. Khodjamirian, R. Ruckl, G. Stoll, and D. Wyler, \plb{402}{1997}{167}; 
Z. Ligeti, L. Randall, and M.B. Wise, \plb{402}{1997}{178}; 
A.K. Grant, A.G. Morgan, S. Nussinov, and R.D. Peccei, \prd{56}{1997}{3151}.
%------------------------
\bibitem{neubert}
A. L. Kagan and M. Neubert, Eur. Phys. J. {\bf C 7} (1999) 5.

\bibitem{gabsarid}
E. Gabrielli and U. Sarid \prd{58}{1998}{115003}; \prl{79}{1997}{4752}.
%------------------------
\bibitem{DTV}
M. A. Diaz, E. Torrente-Lujan, and J.W.F. Valle, \npb{551}{1999}{78}.
%-----------%------------------------

\bibitem{MAR}
A. Czarnecki and W.J. Marciano, \prl{81}{1998}{277}.
%-----------%------------------------
\bibitem{bsgSUSYNLO}
M. Ciuchini, G. Degrassi, P. Gambino, and G.F. Giudice,
\npb{527}{1998}{21};\npb{534}{1998}{3};
F. Borzumati and C. Greub, \prd{58}{1998}{074004};
C. Bobeth, M. Misiak, and J. Urban, \npb{567}{2000}{153}.

\end{thebibliography}
\end{document}